\documentclass[aps,prd,reprint,showpacs,preprintnumbers,bibnotes]{revtex4-1}

\usepackage{graphicx,amsmath,amssymb,hyperref,natbib,slashed}
\usepackage{soul} %For strikethroughs \st{}

\newcommand{\lh}{\textsc{l}}
\newcommand{\rh}{\textsc{r}}

\newcommand{\bea}{\begin{eqnarray}}
\newcommand{\eea}{\end{eqnarray}}
\newcommand{\nn}{\nonumber}
\newcommand{\ov}{\overline}

\newcommand{\rarr}{\rightarrow}

\mathchardef\mhyphen="2D

\def\a{\alpha}

\def\h{\eta}

\def\m{\mu}
\def\n{\nu}

\def\f{\phi}
\def\x{\chi}

\def\D{\Delta}

\def\W{\Omega}

\begin{document}

\title{Searching for Dark Matter at the LHC with a Mono-$Z$}

\author{Nicole F.\ Bell} 
%\email{n.bell@unimelb.edu.au}
\affiliation{ARC Centre of Excellence for Particle Physics at the Terascale, School of Physics, The University of Melbourne, Victoria 3010, Australia}

\author{James B.\ Dent}
\affiliation{Department of Physics and School of Earth and Space Exploration, Arizona State University, Tempe, AZ 85287-1404, USA}
\affiliation{Department of Physics, University of Louisiana at Lafayette, Lafayette, LA 70504-4210, USA}

\author{Ahmad J.\ Galea}
\affiliation{ARC Centre of Excellence for Particle Physics at the Terascale, School of Physics, The University of Melbourne, Victoria 3010, Australia}

\author{Thomas D.\ Jacques}
%\email{tjacques@ph.unimelb.edu.au}
\affiliation{Department of Physics and School of Earth and Space Exploration, Arizona State University, Tempe, AZ 85287-1404, USA}

\author{Lawrence M.\ Krauss}
\affiliation{Department of Physics and School of Earth and Space Exploration, Arizona State University, Tempe, AZ 85287-1404, USA}

\author{Thomas J.\ Weiler}
%\email{t.weiler@vanderbilt.edu}
\affiliation{Department of Physics and Astronomy, Vanderbilt University, Nashville, TN 37235, USA}

\date{\today}

\begin{abstract}
We investigate a mono-$Z$ process as a potential dark matter search
strategy at the LHC.  In this channel a single $Z$ boson recoils
against missing transverse momentum, attributed to dark matter
particles, $\chi$, which escape the detector.  This search strategy is
related, and complementary to, monojet and monophoton searches.  For
illustrative purposes we consider the
process $q\bar{q}\rarr \chi\chi Z$ in a toy dark matter model, where the $Z$ boson is emitted
from either the initial state quarks, or from the internal propagator.
Among the signatures of this process will be a pair of muons with
high $p_T$ that reconstruct to the invariant mass of the $Z$, and large
amounts of missing transverse energy.  Being a purely electroweak
signal, QCD and other Standard Model backgrounds are relatively easily removed
with modest selection cuts.  We compare the signal to Standard Model
backgrounds and
demonstrate that, even for conservative cuts, there exist regions of parameter space where the signal
may be clearly visible above background in future LHC data, allowing either new discovery potential or the possibility of supplementing information about the dark sector beyond that available from other observable channels.
\end{abstract}

\maketitle

%%%%%%%%%%%%%%%%%%%%%%%%%%%%%%%%%%%%%%%%%%%%
\section{Introduction}
%%%%%%%%%%%%%%%%%%%%%%%%%%%%%%%%%%%%%%%%%%%%

There is now compelling cosmological evidence that the dark matter (DM) that appears to dominate galaxies and clusters of galaxies resides in the form of a gas of exotic weakly interacting elementary particles. While this has motivated direct searches for such candidates using underground detectors, and indirect searches attempting to find signatures of dark matter interactions in an astrophysical setting, one of the most exciting possibilities involves producing and detecting such particles  at the LHC.  

Although a multitude of dark matter candidates has been
proposed in the literature (for some reviews see
\cite{Jungman:1995df,DAmico:2009df,Hooper:2009zm,Bergstrom:2012fi,Feng:2010gw,Bertone:2004pz}),
by far the most popular class of model is that of a single massive
particle species which interacts weakly with the standard model
(WIMP).  WIMP models provide an appealing mechanism for dark matter
production in the early Universe, compatible with the observed relic
abundance, and, importantly, allow for the possibility of
direct and indirect detection through a variety of methods today.

Assuming a WIMP-like dark matter candidate $\x$, there are three types
of $\x$-Standard Model (SM) interactions that can be probed
experimentally: annihilation ($\x\x\rarr SM$), scattering ($\x+ SM
\rarr \x + SM$), and production ($SM\rarr \x\x$), which are of interest
for indirect detection, direct detection, and collider searches,
respectively. We consider here the last of these processes,
specifically, the production of $\x$ at the LHC.

If $\x$ couples directly to quarks, it will be produced through quark
annihilation at the LHC, predominantly via the channel $q\bar q
\rarr\x\x$.  Given the stability and weakly interacting nature of the
DM, we expect $\x$ to leave the detectors unseen.  Detection therefore 
requires a visible particle in the final state, against
whose momentum some amount of missing transverse energy
($\slashed{E}_T$) can be reconstructed. The dominant process which
fits this criterion is dependent on the specific model
\cite{Belanger:2012mk}.  In many models, the simplest process that has
both $\x$ and a visible particle in the final state is the
bremsstrahlung of a gauge boson during annihilation, $q\bar{q}\rarr
\chi\chi+gauge\ boson$, where the gauge boson is emitted from
either the initial state quarks or an internal propagator.
Importantly, this process is generic and will therefore always be present, albeit with a rate which is model-dependent.

The lowest-order possibilities for gauge boson bremsstrahlung are the radiation of a single 
gluon (monojet), photon (monophoton), or electroweak gauge boson.
The first two possibilities have been well studied.  The hadronic
monojets subsequent to gluon bremsstrahlung have been considered by
several
groups~\cite{Rajaraman:2011wf,Fox:2011pm,Beltran:2010ww,Shoemaker:2011vi,Aad:2011xw,Chatrchyan:2012fk,Chatrchyan:2012pa,Martinez:2012ie,Hagiwara:2012we,Haisch:2012kf}
and are a promising dark matter search strategy (mono-jets can also be used to probe non-standard neutrino interactions~\cite{Friedland:2011za}).  Similarly,
monophotons ($\gamma$ bremsstrahlung) can be used in dark matter
searches, often in conjunction with jets~\cite{Chatrchyan:2012fk,Malik:2012sa,Aaltonen:2008hh},
although constraints are generally weaker than those derived from
purely jet-based searches~\cite{Fox:2011pm,Frandsen:2012rk}.

The focus of this paper is to investigate collider signatures of dark
matter through purely electroweak bremsstrahlung. These processes can either lead to a mono-$W$~\cite{Bai:2012xg} or mono-$Z$ signal.
Specifically, we consider a mono-$Z$ signal, and highlight some unique kinematical features of this channel, which make it an interesting and important complement to jet and photon based searches.  Because of this unique kinematics, signals may be distinguished from backgrounds even if rates are not as large as for other bremsstrahlung processes.  As a result, depending upon the model, this new signal provides either new discovery potential for dark matter at the LHC, or, equally interesting, information supplemental to other observable channels to further pin down dark matter model-dependent parameters.

We examine the expected signatures of the mono-$Z$
process $q\bar q\rarr\x\x Z$ at the LHC, relative to the SM
backgrounds.  We demonstrate these signatures by implementing a
specific DM model in which the DM couples directly to quarks.  This
is used to demonstrate proof of principle for a
mono-$Z$ dark matter search, rather than being proposed as a fully
self-contained particle physics model.  However, many of the features discussed will be
generically applicable in all WIMP models.  

Electroweak bremsstrahlung has recently received considerable
attention in the context of dark matter annihilation and indirect
detection~\cite{Bell:2010ei,Ciafaloni:2011sa,Bell:2011if,Garny:2011cj,Bell:2011eu,Ciafaloni:2012gs,Garny:2011ii,Ciafaloni:2011gv,Barger:2011jg,Ciafaloni:2010ti,Kachelriess:2009zy,Kachelriess:2007aj,Baro:2011zz,Bell:2012dk}.
In certain models, bremsstrahlung can play an important role in
lifting a helicity suppression of the lowest order annihilation
process, thus becoming the dominant annihilation mode.   
The possibility that lifting helicity suppression might enhance electroweak bremsstrahlung associated with dark matter production at the LHC motivated our initial investigations.
However, we find that helicity unsuppression negligibly affects rates in the kinematically accessible detection regimes which we consider. 
Nevertheless, signals rise above standard model backgrounds.

In Section~\ref{sec:signatures} we describe the dark
matter -- mono-$Z$ signatures at the LHC, and outline the dominant SM
backgrounds.  In Section~\ref{sec:model} we introduce a simple DM
model for which we examine the mono-$Z$ signal, and calculate the
production cross section.  Here we also present a set of kinematic cuts
designed to largely eliminate the SM backgrounds while preserving an
observable number of signal events.  Existing observational
constraints on the model are discussed in
Section~\ref{sec:constraints}, our main results are presented in
Section~\ref{sec:results}, and we conclude in
Section~\ref{sec:conclusions}.

%%%%%%%%%%%%%%%%%%%%%%%%%%%%%%%%%%%%%%%%%%%%
\section{LHC Signatures and Backgrounds}
\label{sec:signatures}
%%%%%%%%%%%%%%%%%%%%%%%%%%%%%%%%%%%%%%%%%%%%

\begin{figure}
\includegraphics[width=0.25\textwidth]{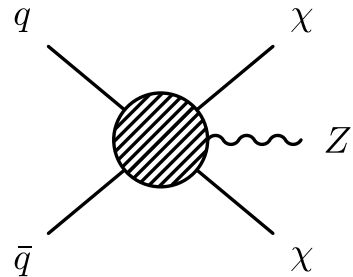}
\caption{Generic electroweak bremsstrahlung process, $q\bar q\rarr\x\x
  Z$, which leads to a mono-$Z$ signal at the LHC.\label{fig:brem-blob}}
\end{figure}

The $\chi\chi Z$ production process is pictured schematically in
Fig.~\ref{fig:brem-blob}.  In Section~\ref{sec:model} we will
introduce a specific DM model in order to explore this process in more detail.
Here we will outline some of the general features of the
$\chi\chi Z$ final state, along with the relevant SM backgrounds.

%%%%%%%%%%%%%%%%%%%%%%%%%%%%%%%%%%%%%%%%%%%%
\subsection{The $Z\chi\chi$ final state}
\label{subsec:zchichi}
%%%%%%%%%%%%%%%%%%%%%%%%%%%%%%%%%%%%%%%%%%%%

Key to the discovery of the $\chi\chi\,(Z\rightarrow f{\bar f})$ final state from within the myriad of 
SM backgrounds is the correct reconstruction of a $Z$ boson from the
invariant mass of its decay products.  We consider the muonic decay
mode which, while having a low branching fraction ($\sim 3\%$),
provides for a very clean invariant mass reconstruction.  It also has
the benefit of having few backgrounds relative to hadronic decay
modes.

%%%%%%%%%%%%%%%%%%%%%%%%%%%%%%%%%%%%%%%%%%%%
%%%%%%%%%%%%%%%%%%%%%%%%%%%%%%%%%%%%%%%%%%%%

The recoil of the $Z$ against the heavy dark matter particles results in 
decay muons with large transverse momentum.  
The signal of interest will therefore be a pair
of high $p_T$ muons, with an invariant mass that reconstructs to that
of the $Z$, and a momentum sum which reveals a large amount of
$\slashed{E}_T$.

We simulate both the signal and background process in the MadEvent~\cite{Alwall:2011uj} event generator.  The MadEvent output is then fed into
PYTHIA~\cite{Sjostrand:2007gs} to simulate higher order initial state
and propagator radiation effects.  A detailed detector simulation is
beyond the scope of this work, as our goal is to demonstrate the
potential for mono-$Z$ dark matter processes to be observed above
background at the LHC, rather than to calculate precise constraints on
specific dark matter models.

%%%%%%%%%%%%%%%%%%%%%%%%%%%%%%%%%%%%%%%%%%%%
\subsection{Backgrounds}
\label{subsec:Backgrounds}
%%%%%%%%%%%%%%%%%%%%%%%%%%%%%%%%%%%%%%%%%%%%
 
The backgrounds for our process come from channels producing a dimuon pair and $\slashed{E}_T$. The dominant backgrounds are the leptonic decays of gauge boson producing processes, specifically $ZZ$, $ZW^\pm$, $W^+ W^-$ and $t\ov{t}\rarr b\ov{b}W^+ W^-$. 

Production of $Z+jets$ can also contribute a $\slashed{E}_T$
background through jet mismeasurement in the hadronic calorimeter.
To test the importance of this background, we used MadEvent to
simulate the dominant contribution, $Z$ plus a single jet.
Hadronization was performed in PYTHIA, and fast detector simulation was carried out (for $Z+jet$ background alone) in Delphes~\cite{Ovyn:2009tx} (using ATLAS parameters), which simulates calorimeter smearing and reconstructs $\slashed{E}_T$. As expected, this background was found to be sub-dominant to the other backgrounds after the cuts described in Section~\ref{subsec:EventSelection}. At leading order, and in the absence of full detector simulation and pile-up effects, the accuracy of our treatment of this background is obviously limited.  Our simulations should, however, be accurate to within an order of magnitude of the true background, which is adequate for our purposes.

As in the $Z+jet$ case, mismeasurement of $b$ jets will contribute to the $\slashed{E}_T$ spectrum for the process $t\ov{t}\rarr b\ov{b}W^+ W^-$. This contribution is expected to be small compared to neutrinos from $W^{\pm}$ decay, and is therefore neglected.  Similarly, mismeasurement of initial state radiation (ISR) in the form of gluon jets can contribue a $\slashed{E}_T$ background.
Given the limited accuracy of PYTHIA in simulating these higher order process, we do not consider these effects.  However, we expect their contributions to $\slashed{E}_T$ will be small, based on our simulations of the $Z+jet$ background.

\begin{figure}
\includegraphics[width=0.5\textwidth]{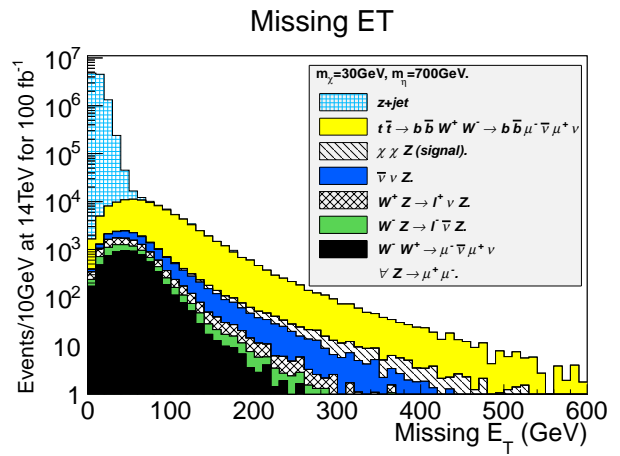}
\caption{Modes contributing to full $\slashed{E}_T$ spectrum for signal and background 
for $m_\x=30$GeV and $m_\eta=700$GeV, 
after inclusive $p_T$ and invariant mass cuts. 
The signal cross section ($\rarr \chi\chi\,Z$) is calculated with a coupling that produces the observed relic abundance.\label{fig:FullET}}
\end{figure}

The relative contributions of various backgrounds before the implementation of the full set of cuts employed in this work can be seen in Fig.~\ref{fig:FullET}. Cuts on the invariant mass  of the muon pairs ensure that NLO contibutions from $\gamma\rarr\mu^{+}\mu^{-}$ are negligible in these processes. 
The NNLO process $gg\rarr ZZ$ can modify the $ZZ$ background by up to 15\% \cite{Binoth:2008pr}; 
given the level of accuracy desired in this work, we neglect these contributions.
These backgrounds are further reduced or eliminated through cuts described in Section~\ref{subsec:EventSelection}. 

As evident from Fig.~\ref{fig:FullET}, the $Z+jet$ background is
substantial.  This same final state is also the dominant background for the related
monojet DM search channel.  However, we expect this background can be
removed more easily for mono-$Z$'s than for monojets.  For a monojet
search the invisible decays $Z\rarr \nu\bar\nu$ 
provide a large $\slashed{E}_T$ background, with kinematics very
similar to the $\chi\chi + jet$ signal searched for.  In contrast,
$Z+jets$ contributes to the mono-$Z$ background through $Z+jet \rarr
\m^+\m^- +jet$, with $\slashed{E}_T$ arising only via
jet mismeasurement.  This is kinematically very diferent from our
$\chi\chi+Z$ signal and, as we will show below, can be removed
relatively easily with selection cuts.  The sub-dominance of high
cross section QCD backgrounds relative to electroweak processes is an
appealing aspect of the mono-$Z$ signal.

%%%%%%%%%%%%%%%%%%%%%
%%%%%%%%%%%%%%%%%%%%

%%%%%%%%%%%%%%%%%%%%%%%%%%%%%%%%%%%%%%%%%%%%
\section{The Model and Event Selection}
\label{sec:model}
%%%%%%%%%%%%%%%%%%%%%%%%%%%%%%%%%%%%%%%%%%%%

To illustrate the potential for observing a mono-$Z$ dark matter signal
at the LHC, we introduce a toy model in which this process has a
significant rate.  We will then detail event selection criteria that
allow the backgrounds to be largely removed.

%%%%%%%%%%%%%%%%%%%%%%%%%%%%%%%%%%%%%%%%%%%%
\subsection{An Example DM Model}
\label{subsec:ToyModel}
%%%%%%%%%%%%%%%%%%%%%%%%%%%%%%%%%%%%%%%%%%%%

We take the DM to be a gauge-singlet Majorana fermion, $\x$, which
couples to the quark doublet, $Q_\lh$, via the interaction term
 \bea
\mathcal{L}_\text{int}&=&f_{ud}\bar Q
_\lh\h\x_\rh+h.c\nn\\ &=&f_{ud}\left(\h_u\ov{u}_\lh+\h_d\ov{d}_\lh\right)\x_\rh+h.c.,\label{eq:lag}
\eea 
where $f_{ud}$ is a coupling constant and $\h$ is a scalar
field that that transforms as $\h\sim (3,2,1/3)$ under the SM gauge
groups.  (This model is a related to that of ref.~\cite{Cao:2009yy},
modified such that the scalar is charged under $SU(3)_C$.)

Such couplings are also present in supersymmetric (SUSY) models, with
$\x$ identified as a neutralino and $\h$ a squark doublet.  An obvious
difference, however, is that we have no gluino analogue in our model.
In some sense this model is analogous to a SUSY model in which the
gluinos are too heavy to be kinematically accessible at the LHC.

As a consequence, despite this model being substantially simpler than many SUSY models, both in couplings and free parameters, the LHC signatures presented in this work may still be of relevance for some SUSY searches (especially if the parameter space of more minimal SUSY models is increasingly ruled out), 
perhaps providing a complementary signal to further constrain models.
 
\begin{figure*}
\includegraphics[width=0.2\textwidth]{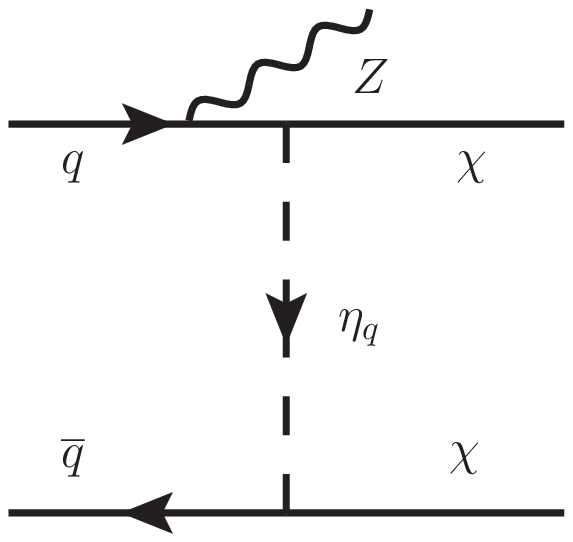}
%\hspace{1cm}
\includegraphics[width=0.2\textwidth]{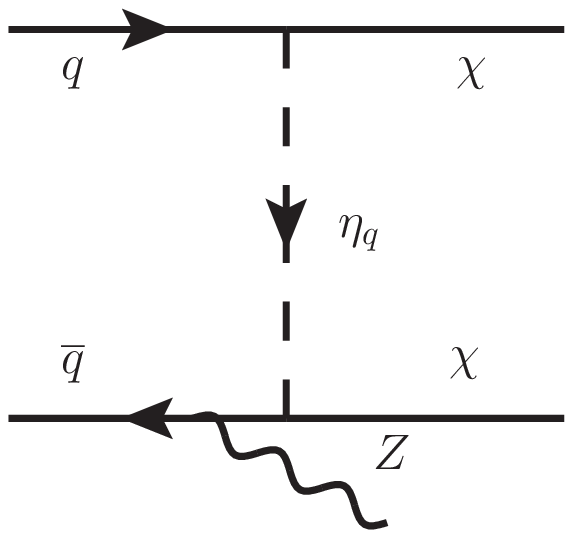}
%\hspace{1cm}
\includegraphics[width=0.2\textwidth]{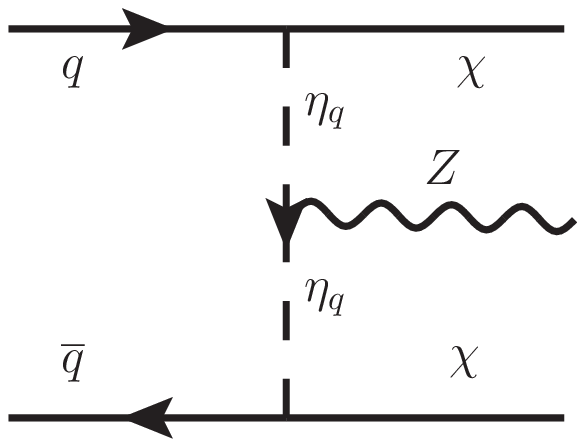}
\caption{t-channel processes contributing to electroweak bremsstrahlung in annihilations to dark matter.
Not shown are the three corresponding u-channel diagrams.
\label{fig:AnDiag}}
\end{figure*}

The interactions in Eq.~\ref{eq:lag} allow for direct annihilation of quarks into $\x$ pair via $t$-channel and $u$-channel $\h$ exchange. Of interest to this work are processes to the next order in $\a_W$, in which a $Z$ boson is radiated from the initial state quarks or the internal propagator. Contributing to the mono-$Z$ process $q\bar q \rarr\x\x Z$ are the three $t$- and $u$-channel diagrams shown in Fig.~\ref{fig:AnDiag}.

%%%%%%%%%%%%%%%%%%%%%%
%%%%%%%%%%%%%%%%%%%%%%

%%%%%%%%%%%%%%%%%%%%%%
\begin{figure}[h!]
\includegraphics[width=0.5\textwidth]{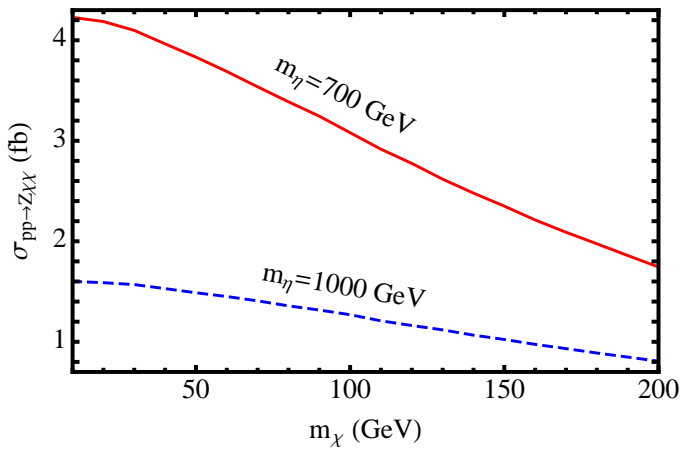}
\caption{ Cross section for process $p p \rarr \chi \chi Z$ at CoM of 14TeV as a function of DM mass. Red line (upper) corresponds to $m_\eta=700$GeV, blue (lower) to $m_\eta=1$TeV. Both cross sections calculated for $f_{ud}=1$, and using CTEQ6L1 PDF's.\label{fig:CrossSec}}
\end{figure} 
%%%%%%%%%%%%%%%%%%%%%%

The Mathematica package FeynRules~\cite{Christensen:2008py} was used to formulate the Feynman rules for the model of Eq.~\ref{eq:lag}.
The rules were interfaced with the MadGraph package~\cite{Alwall:2011uj} to calculate the scattering amplitudes for the processes in Fig.~\ref{fig:AnDiag}. These amplitudes were then input into the MadEvent event generator, which calculated the LHC-relevant cross section $pp\rarr\x\x Z$ for a given $f_{ud}$, 
and for various choices of $m_\x$ and $m_\h$ . 
All cross section calculations were performed in the proton center of momentum frame. 
The probability amplitudes were integrated over the CTEQ6L1~\cite{Lai:2010nw} parton distribution functions (PDF). Given that the LHC is a proton-proton collider, the valence quarks are all $q$ distributions, while the sea-quarks
are of course $q$ and $\bar q $ distributions.

Cross sections at 14TeV Centre of Mass (CoM) are displayed in Fig.~\ref{fig:CrossSec} as a function of $m_\x$, for values of $m_\eta$ relevant to electroweak-scale physics.

%%%%%%%%%%%%%%%%%%%%%%%%%%%%%%%%%%%%%%%%%%%%
\subsection{Event Selection}
\label{subsec:EventSelection}
%%%%%%%%%%%%%%%%%%%%%%%%%%%%%%%%%%%%%%%%%%%%

Now that we have a model which produces DM along with a $Z$-boson, we will examine how this channel may be detected at a hadron collider.  While the backgrounds presented in ~\ref{subsec:Backgrounds} have rates much larger than our signal, they can be substantially reduced with an educated set of cuts on measured events. 
We make cuts on the invariant mass of the muon pair within a 60GeV window 
centered on
the $Z$ mass, which greatly reduces the contribution from non-$Z$ backgrounds, namely $W^+W^-$ and $t\ov{t}$ production.

The presence of the heavy $\x$ in our signal process ensures large amounts of $\slashed{E}_T$. This can be seen clearly in Fig.~\ref{fig:FullET},
which shows the number of expected collider events as a function of missing energy, in 10GeV bins. 
As expected, the number of signal events with large $\slashed{E}_T$ are at least comparable to all SM backgrounds. 
We choose a missing energy cut of $\slashed{E}_T>150$GeV to remove a large fraction of the background events, including the bulk of the $Z+jet$ background. It is important to note that due to the very large cross sections for $t\ov{t}$ and $Z+jet$ before the implementation of cuts, the statistics in these two contributions lead to evident 
fluctuations at high $\slashed{E}_T$ in Fig.~\ref{fig:FullET}.

The $Z$ in the final state can be highly boosted by its recoil off the heavy DM particles; 
we therefore expect its decay products to have large $p_T$. 
We apply the conservative inclusive cut of $p_T>50$GeV on the muon transverse momentum (i.e. require at least one muon in final state with $p_T>50$GeV).

A further consequence of the $Z$ being produced relativistically is that the muons from the decay process will be produced nearly co-linear with each other. This co-linearity ensures a low $\D R$ 
between the pair, where $\D R$ is defined to be
\bea
\D R \equiv\sqrt{\D \f^2 + \D \h^2},
\eea 
where $\phi$ is the azimuthal angle and $\eta$ is the pseudo-rapidity of a particle in the detector.
 
\begin{figure}
\includegraphics[width=0.5\textwidth]{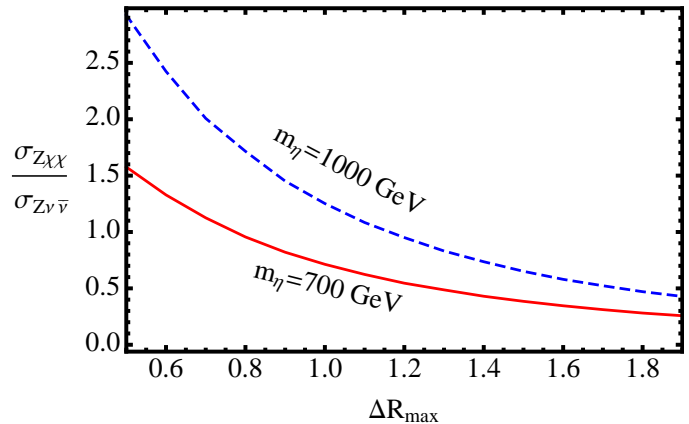}
\caption{ Ratio $\sigma_{\x\x Z}/\sigma_{Z\n\bar\n}$ at 14TeV CoM and for $m_\x=30$ GeV, as a function of the cut on maximum $\Delta R = \sqrt{ \Delta \phi^2 + \D \eta^2}$ between muons in the final state. Red line (lower) corresponds to $m_\eta=700$GeV, blue (upper) to $m_\eta=1$TeV.\label{fig:RDeltaR}}
\end{figure}

Figure \ref{fig:RDeltaR} shows the ratio of signal cross section to the $\nu\bar\nu\, (Z\rarr\m^+\m^-)$ background as a function of cut on maximum $\D R$, after $p_T$ and dimuon invariant mass cuts, for selective points in the model parameter space. This background is useful for comparisons, as it is the dominant $\slashed{E}_T$ background in the region of interest. The signal to background is maximized for lower $\D R$, with both cross-sections becoming equal around $\D R_{max}\sim1$. To preserve signal events, we choose the conservative cut of $\D R<1$. This cut should effectively discriminate against the $W^+W^-$ and $t\ov{t}$ backgrounds, which produce muon pairs with a broad $\D R$ distribution. The effectiveness of this choice of cut can be seen in the top right hand panel of Fig.~\ref{fig:results-meta}, which displays the same data as Fig.~\ref{fig:FullET} 
but with full set of cuts applied, including the cut on $\D R$.

The missing energy in the background $Z+jet$ is a result of jet mismeasurement in the hadronic calorimeter, thus a large amount of $\slashed{E}_T$ is present in events with high jet $p_T$. These types of events typically have a highly boosted $Z$, and subsequently produce muon pairs with 
lower $\D R$ separation than low $\slashed{E}_T$ events. 

We note that the detector simulation program
Delphes requires that muons counted individually be isolated within a cone of $\D R<0.5$.
A consequence of these two factors is that in the majority of $Z+jet$ events with $\slashed{E}_T>150$GeV, muons from $Z$ decay do not pass Delphes' isolation criterion and are subsequently rejected, reducing this background significantly. 
Aside from necessarily using Delphes for producing the $\slashed{E}_T$ spectrum of the $Z+jet$ background, we do not use detector simulation in this work.
Instead, 
we enforce $\D R >0.5$ between the paired muons.

Also, in this work we have been conservative in that we have not applied any jet related cuts, despite their obvious utility in distinguishing real $\slashed{E}_T$ from jet mismeasurment. 
It is possible to further improve signal relative to QCD backgrounds by removing events 
in which jet and $\slashed{E}_T$ directions are correlated. 
Tagging of $b$ jets may also be used to reduce the $t\ov{t}$ background.
Below we show that such additional cuts, while no doubt useful, 
are not necessary to obtain an observable signal to background ratio.

%%%%%%%%%%%%%%%%%%%%%%
%%%%%%%%%%%%%%%%%%%%%%

%%%%%%%%%%%%%%%%%%%%%%%%%%%%%%%%%%%%%%%%%%%%
\section{Model Constraints}
\label{sec:constraints}
%%%%%%%%%%%%%%%%%%%%%%%%%%%%%%%%%%%%%%%%%%%%

When calculating LHC signals, we adopt model parameters that produce
the correct DM relic abundance.  We also ensure the parameters are in
accordance with current direct detection and collider bounds.  These
constraints are described in detail here.

%%%%%%%%%%%%%%%%%%%%%%%%%%%%%%%%%%%%%%%%%%%%
\subsection{Freezeout\label{subsec:freezeout}}
%%%%%%%%%%%%%%%%%%%%%%%%%%%%%%%%%%%%%%%%%%%%

We work in the context of the standard WIMP scenario, in which the DM
was in thermal equilibrium in the early Universe up until the
time of thermal freezeout, at which point the relic abundance was set.
For a given DM mass, we wish to choose values of the coupling constant
$f_{ud}$, and $\eta$ mass, such that
the DM freezes out with the correct relic abundance. 

\begin{figure}
\includegraphics[width=0.45\textwidth]{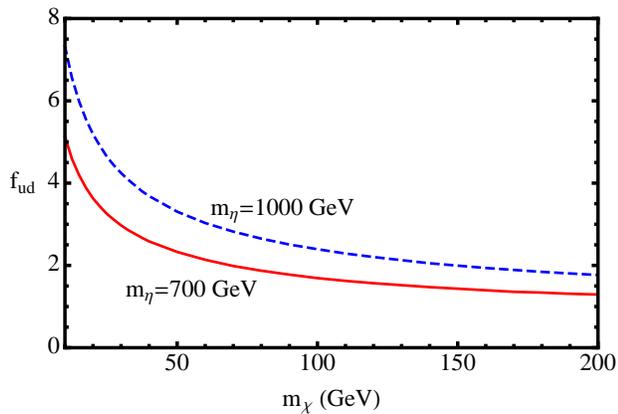}
\caption{The minimum allowed coupling constant $f_{ud}$ in order to satisfy the requirement 
that the contribution to the DM relic density from this model be less than or equal to the total DM relic density, 
$\W_{DM} h^2 \simeq 0.11$. Red line (lower) corresponds to $m_\eta=700$GeV, blue (upper) to $m_\eta=1$TeV. 
Note that the expansion parameter $f^2/4\pi$ remains perturbative for $f\alt 4\pi\sim 12.5$.
\label{fig:fconstraints}}
\end{figure}

The process which kept $\x$ in equilibrium before
thermal freezeout was $q\bar q \rarr\x\x$. The relic density of $\x$
was therefore determined by parameters $f_{ud}$, $m_\x$ and $m_\eta$. Following
\cite{Steigman:2012nb,Gondolo:1990dk}, we use a semi-analytic
solution to the co-moving Boltzmann equation, and the inferred value
$\W_{DM} h^2\simeq0.11$ to place constraints on $f_{ud}$ for given
values of $m_\x$ and $m_\eta$.
Results are displayed in Fig.~\ref{fig:fconstraints}.
If the coupling were any smaller than the constraints in Fig.~\ref{fig:fconstraints}, then the 
DM would have been overproduced in the early universe, yielding an abundance greater than that observed today. 
On the other hand, if it were any larger, then the relic abundance would be smaller than observed. 
If there are other DM species contributing to the relic abundance, then the constraints on $f_{ud}$ serve as lower limits, since the DM candidate under consideration need not contribute the entire relic abundance.

%%%%%%%%%%%%%%%%%%%%%%
%%%%%%%%%%%%%%%%%%%%%%

%%%%%%%%%%%%%%%%%%%%%%%%%%%%%%%%%%%%%%%%%%%%
\subsection{Direct Detection}
%%%%%%%%%%%%%%%%%%%%%%%%%%%%%%%%%%%%%%%%%%%%

\begin{figure}
\includegraphics[width=0.2\textwidth]{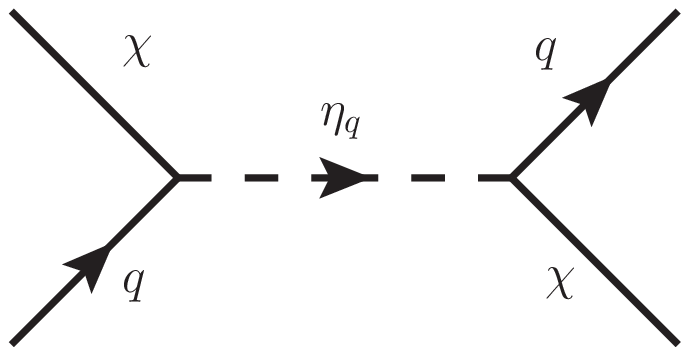}
\includegraphics[width=0.2\textwidth]{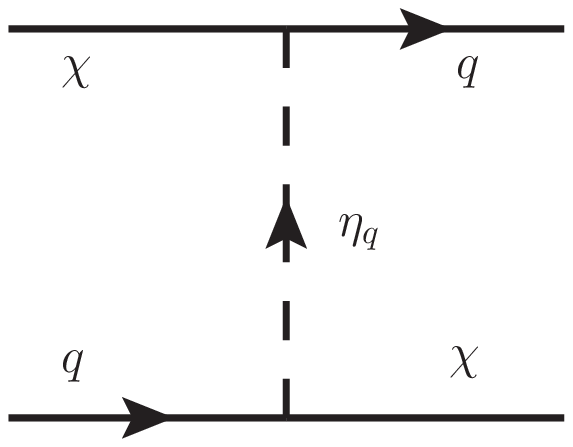}
\caption{Fundamental processes contributing to DM--nucleon scattering.\label{fig:DDProc}}
\end{figure}

In the model under discussion, quarks couple to the beyond-SM sector
via a $q\chi\eta$~vertex with strength $f_{ud}$.  Consequently, care
is required to avoid direct detection constraints.  The operator in
Eq.~\ref{eq:lag} allows for $\chi$-quark scattering via the $s$ and
$u$-channel $\h$ exchange graphs in Fig.~\ref{fig:DDProc}, which can
in turn be related to $\x$-nucleon scattering.

%%%%%%%%%%%%%%%%%%%%%%
%%%%%%%%%%%%%%%%%%%%%%

\begin{figure}
\includegraphics[width=0.45\textwidth]{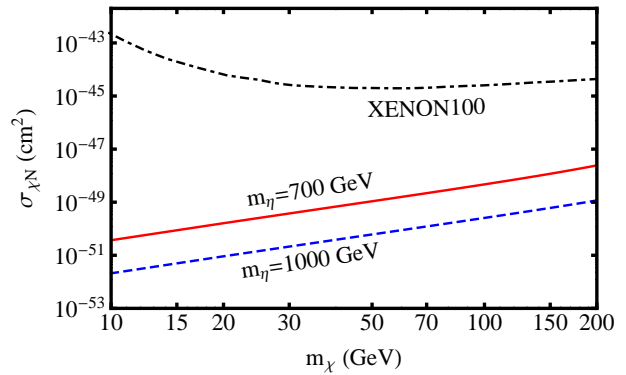}
\caption{The SI $\chi -$nucleon scattering cross section. Red line (upper) corresponds to $m_\eta=700$GeV, blue (lower) to $m_\eta=1$TeV. Shown in dashed is the upper limit on the SI cross section from XENON100 experiment. \label{fig:DDconstraints}}
\end{figure}

The couplings in Eq.~\ref{eq:lag} Fierz transform into couplings to nucleons that have both spin-dependent (SD) and spin-independent (SI) contributions. 
The strongest constraints on our model come from the spin-independent limits from the XENON100 experiment \cite{Baudis:2012zs}, 
which looks for excitations in ultra-cold liquid Xe resulting from DM scattering events. 
We performed the calculation of the SI $\x$-nucleon cross section in the current model using the micrOMEGAs~\cite{Belanger:2010gh} software package, 
taking the Lagrangian in Eq.~\ref{eq:lag} as input. The calculation was done for  values of $f_{ud}$ that produce 
the correct relic abundance (Fig.~\ref{fig:fconstraints}) for various values of $m_\eta$ and a range of DM masses. 
The results are displayed in Fig.~\ref{fig:DDconstraints}, alongside the upper bound on the cross section allowed by XENON100. Clearly the model parameters considered in this work are allowed by the XENON100 constraint.
Note that if there are additional DM particles, and the DM candidate considered here is not required to contribute the entire relic density, the curves in Fig.~\ref{fig:DDconstraints} denote lower limits on the scattering cross section.

%%%%%%%%%%%%%%%%%%%%%%
%%%%%%%%%%%%%%%%%%%%%%

%%%%%%%%%%%%%%%%%%%%%%%%%%%%%%%%%%%%%%%%%%%%
\subsection{Collider Constraints}
%%%%%%%%%%%%%%%%%%%%%%%%%%%%%%%%%%%%%%%%%%%%

Through the early part of 2012, roughly 5 fb$^{-1}$ of data have been analyzed by both the ATLAS
and CMS collaborations.  Thus far no significant discrepancies with Standard Model predictions have been found.
Discrepancies not seen at present include large missing energy signals and new particles such as 
those predicted in numerous extensions of the SM.

The absence of novel signals has put ever tighter constraints on
models of physics beyond the SM.  As discussed in
Section~\ref{subsec:ToyModel}, the model described in the present work
is purely phenomenological, but does possess some similarities to a
SUSY model (with obvious differences being an absence of gluinos and
only first generation quark couplings). This makes it somewhat
difficult to directly compare existing bounds with our model.
However, we have chosen values of parameters such as the $\eta$
masses and $\eta-\chi$ mass differences which are not currently
ruled out by squark mass and squark-neutralino mass difference
constraints, respectively, for a simplified model spectra (SMS) of SUSY.

Examples of 
these constraints for an SMS from CMS data are given in ~\cite{CMS-PAS-SUS-11-016, Kribs:2012gx, Dreiner:2012gx, Beskidt:2012sk}.  The SMS are motivated by popular SUSY frameworks such as the constrained minimal supersymmetric standard model (cMSSM) and the general gauge mediation model (GGM). Of particular relevance are the constraints derived in ~\cite{Kribs:2012gx}, in which the authors consider a simplified model with a heavy gluino and two squark generations. This very closely resembles the model at hand, the only effective difference being the number of scalar degrees of freedom present in the model. A lower bound of 780GeV was placed on the squark mass, which maps to the constraint $m_\eta \gtrsim 600$GeV in the model considered here. 
Conservatively, we only consider $\eta$ masses above 700GeV.

%%%%%%%%%%%%%%%%%%%%%%%%%%%%%%%%%%%%%%%%%%%%
\section{Results}
\label{sec:results}
%%%%%%%%%%%%%%%%%%%%%%%%%%%%%%%%%%%%%%%%%%%%

\begin{figure*}[!htbp]
\includegraphics[width=0.47\textwidth]{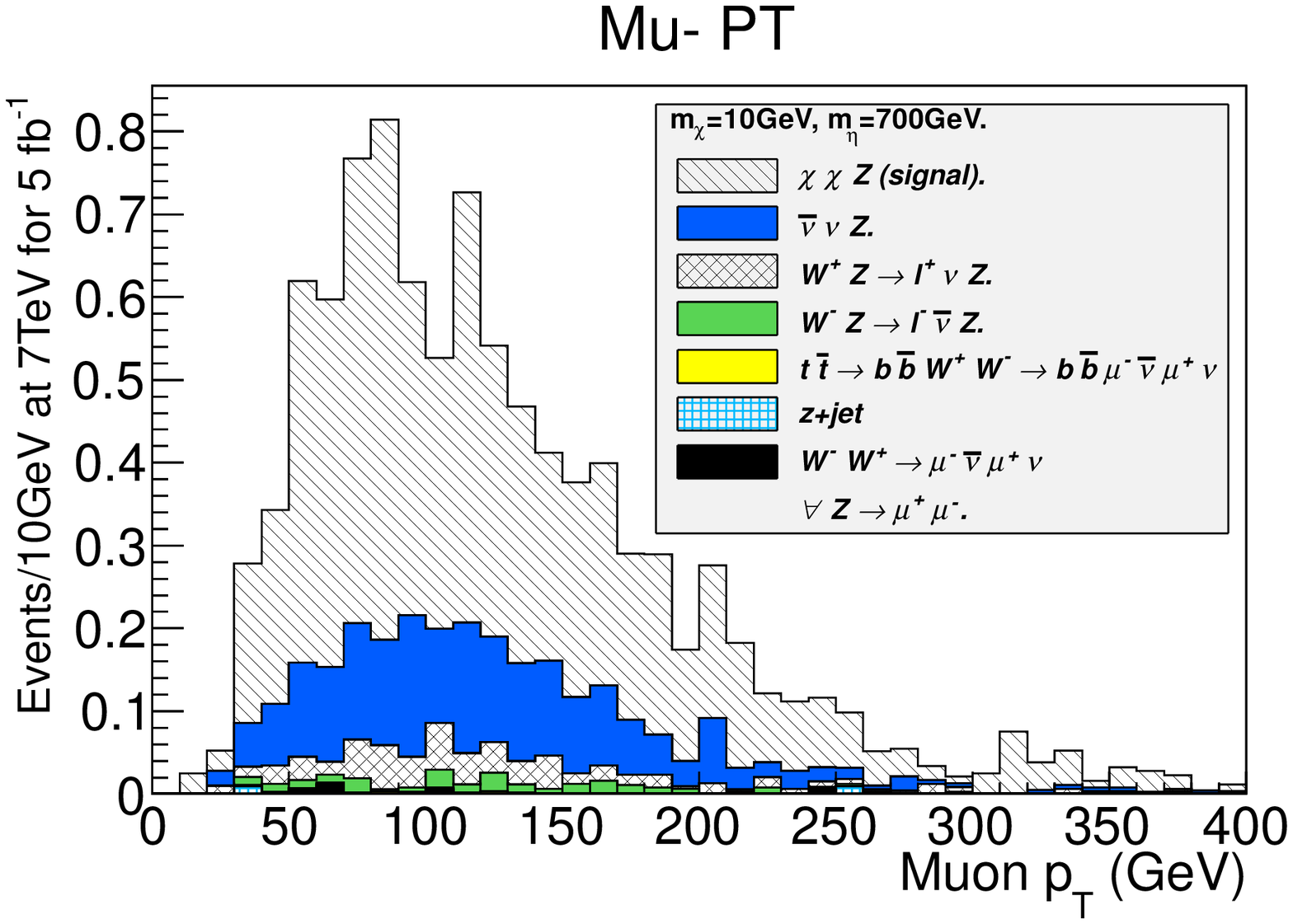}
\includegraphics[width=0.47\textwidth]{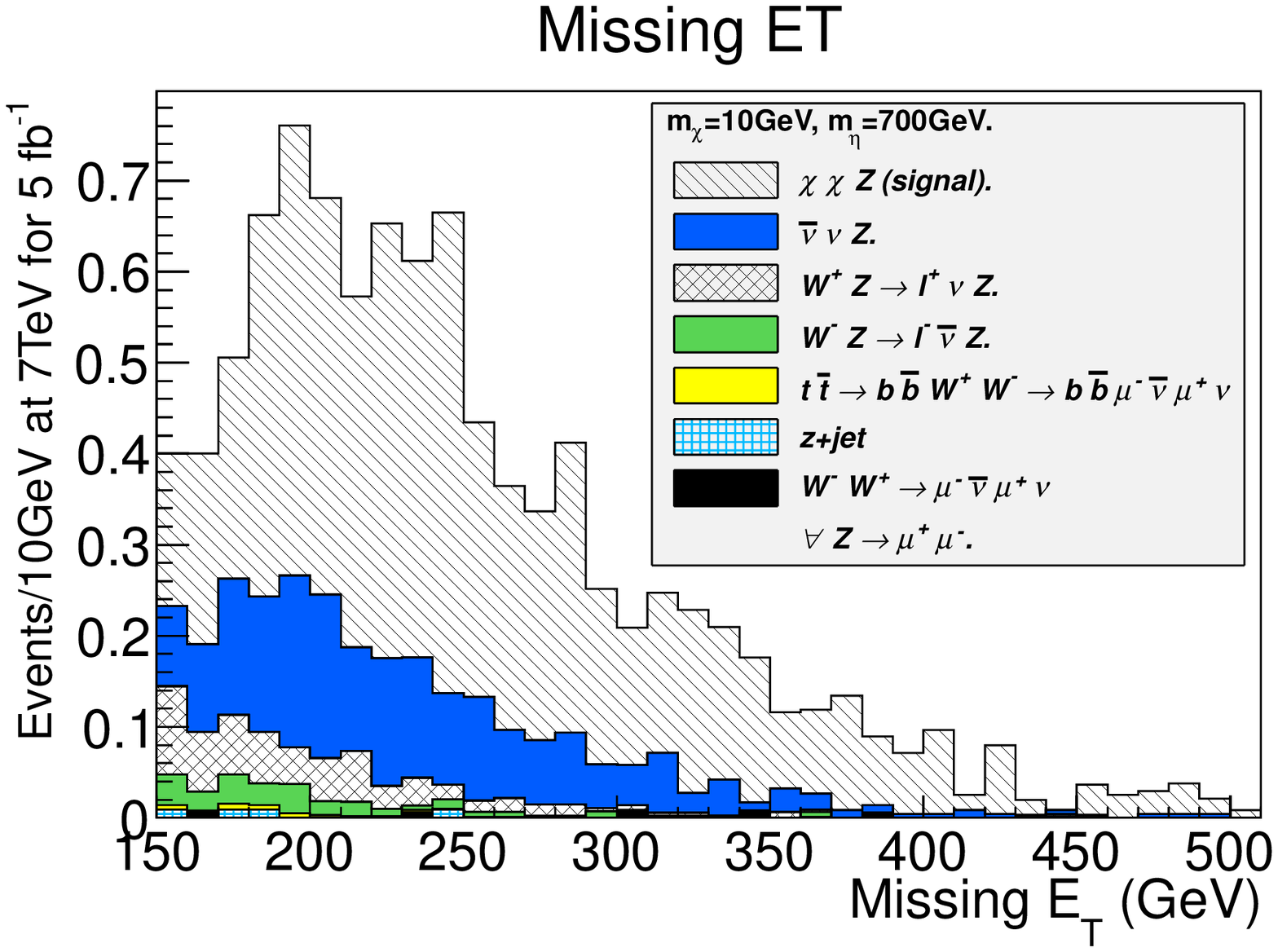}
\includegraphics[width=0.47\textwidth]{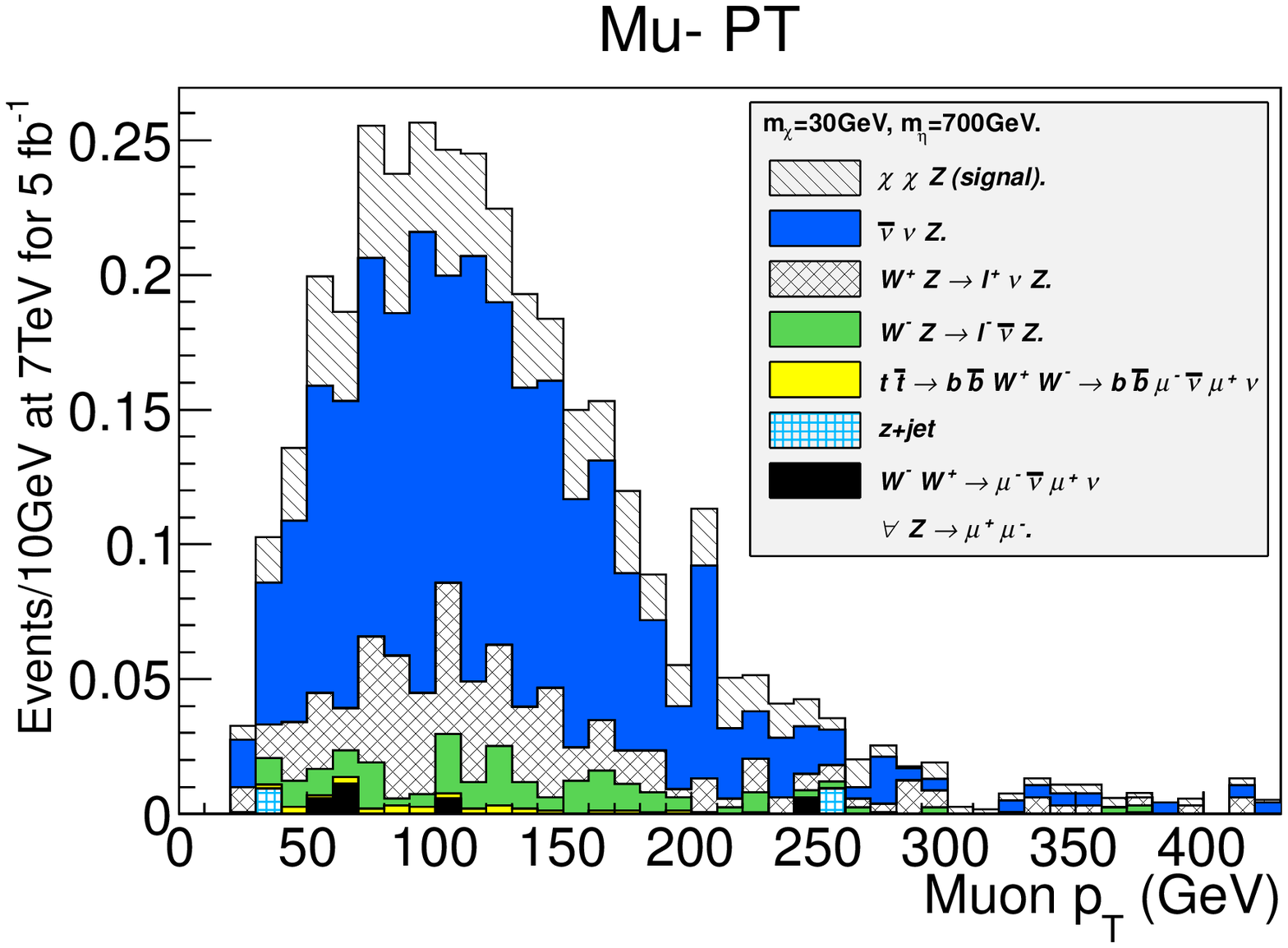}
\includegraphics[width=0.47\textwidth]{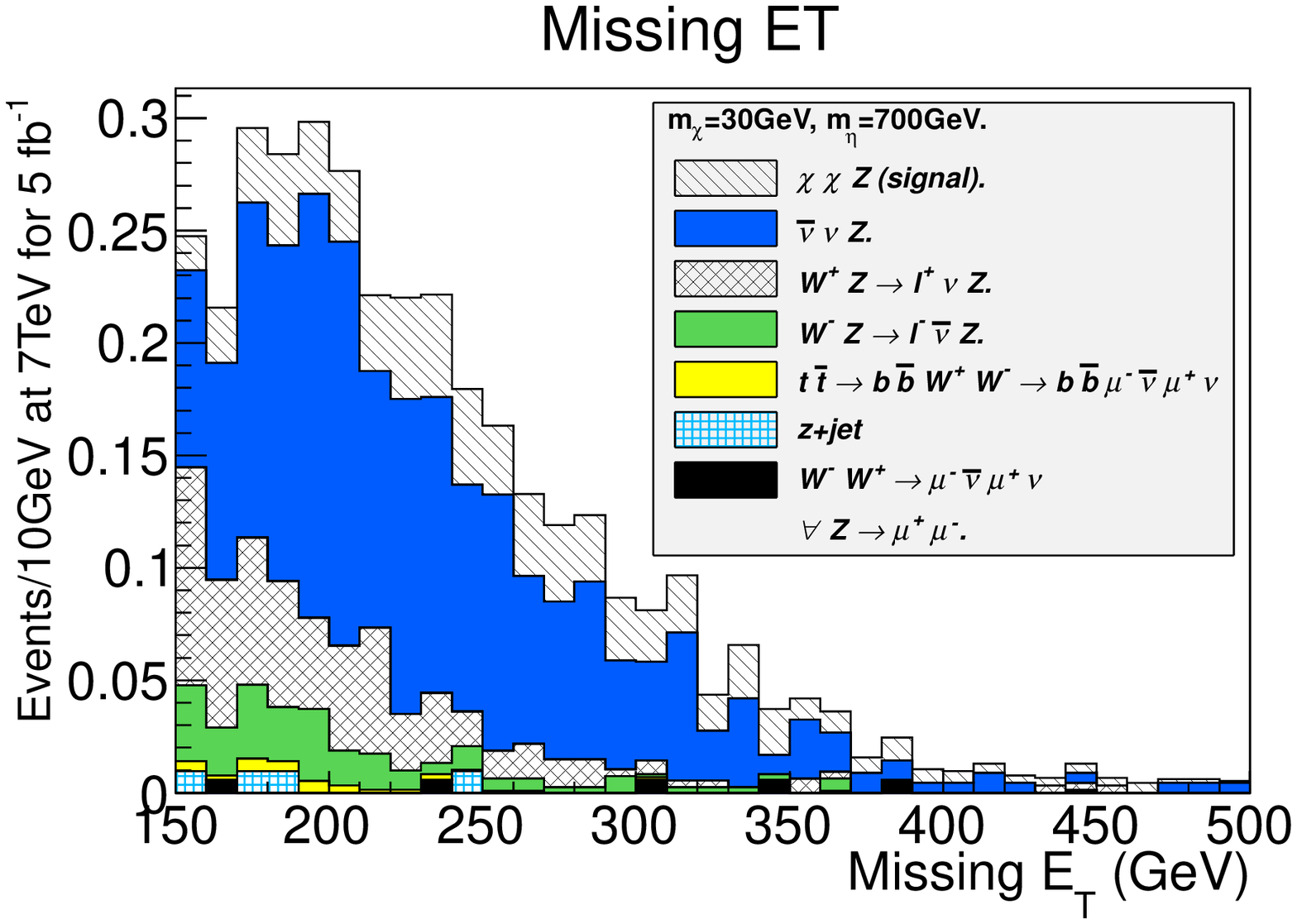}
\caption{Events passing selection criteria in Section~\ref{subsec:EventSelection}, at $\sqrt{s}=7$TeV and 5$fb^{-1}$ of data, for $m_\eta=700$GeV, comparing  $m_\x=10$GeV and 30GeV.\label{fig:results-7TeV}}
\end{figure*}

\begin{figure*}[!htbp]
\includegraphics[width=0.47\textwidth]{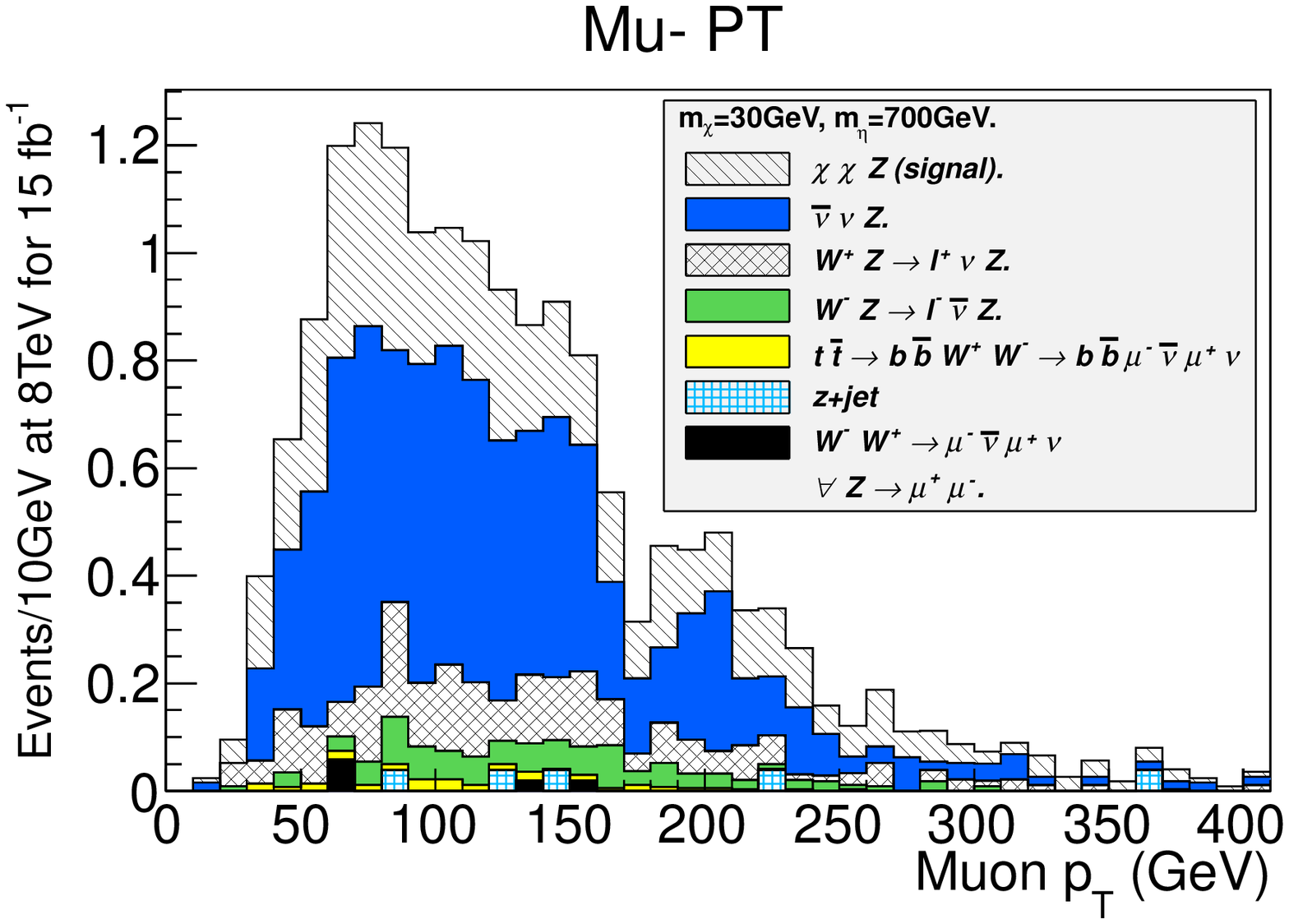}
\includegraphics[width=0.47\textwidth]{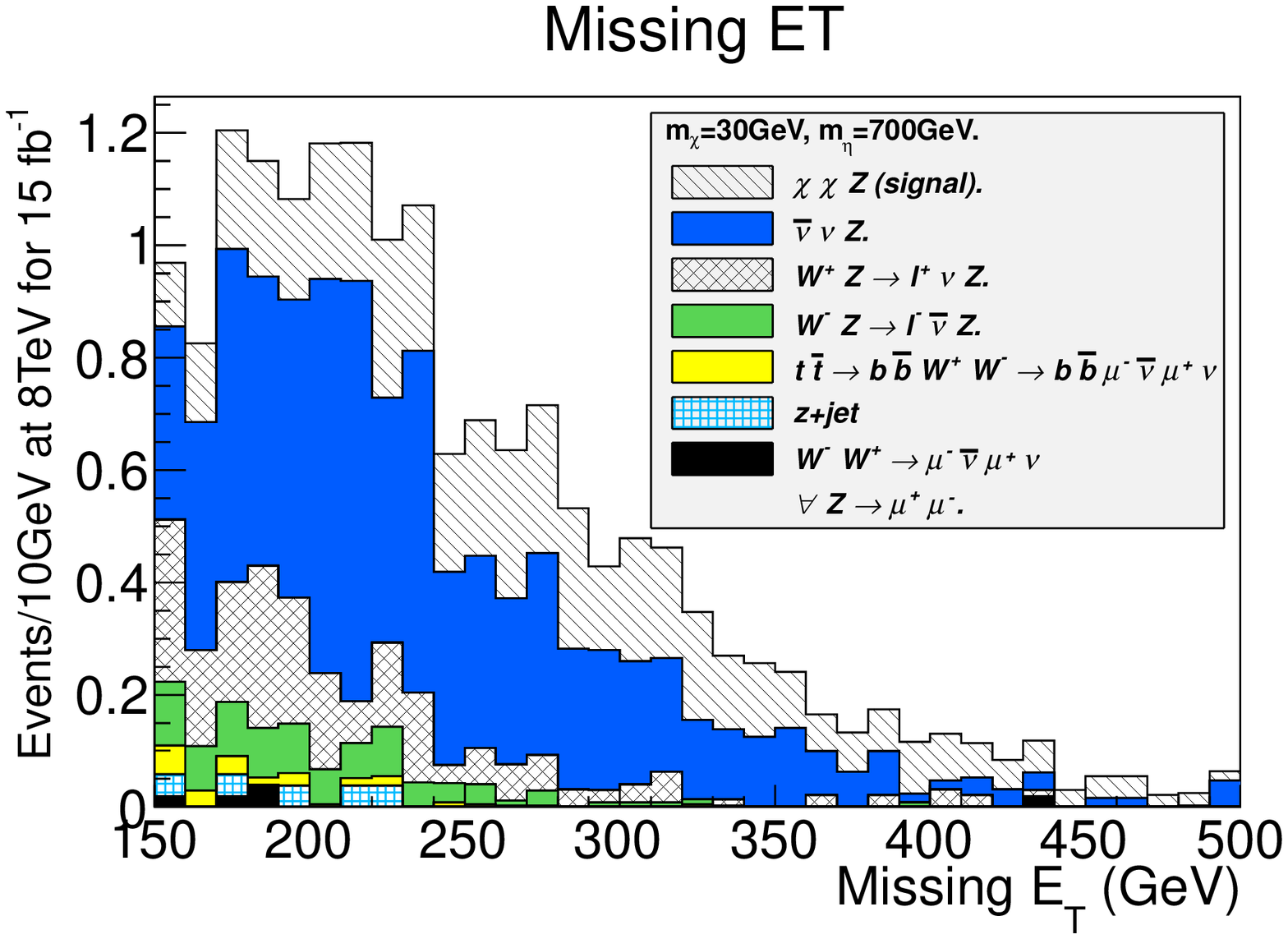}
\caption{As Fig~\ref{fig:results-7TeV}, with $\sqrt{s}=8$TeV and 15$\text{fb}^{-1}$ of data, for $m_\x=30$GeV and $m_\eta=700$GeV.\label{fig:results-8TeV}}
\end{figure*}

\begin{figure*}[!htbp]
\includegraphics[width=0.44\textwidth]{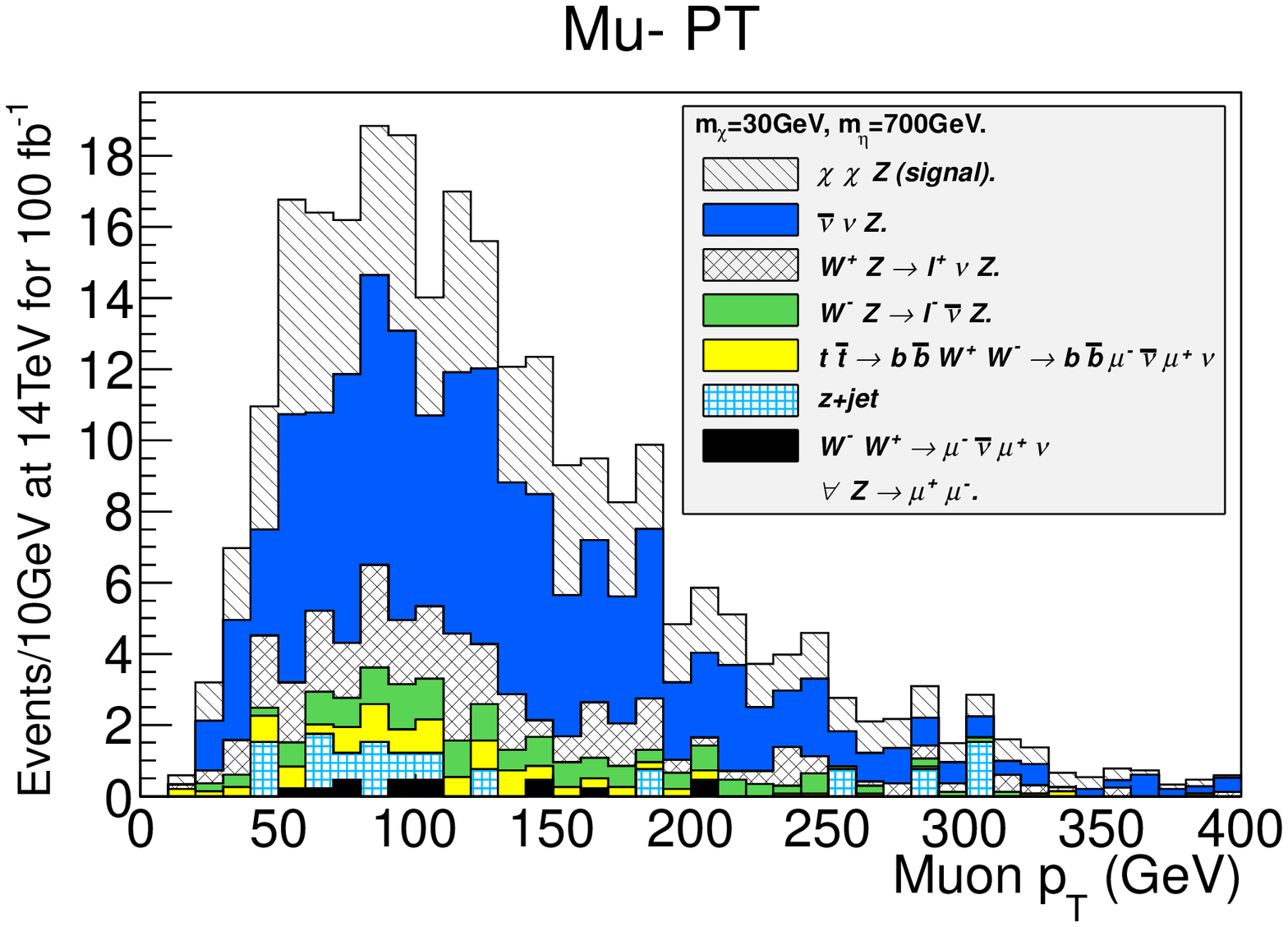}
\includegraphics[width=0.44\textwidth]{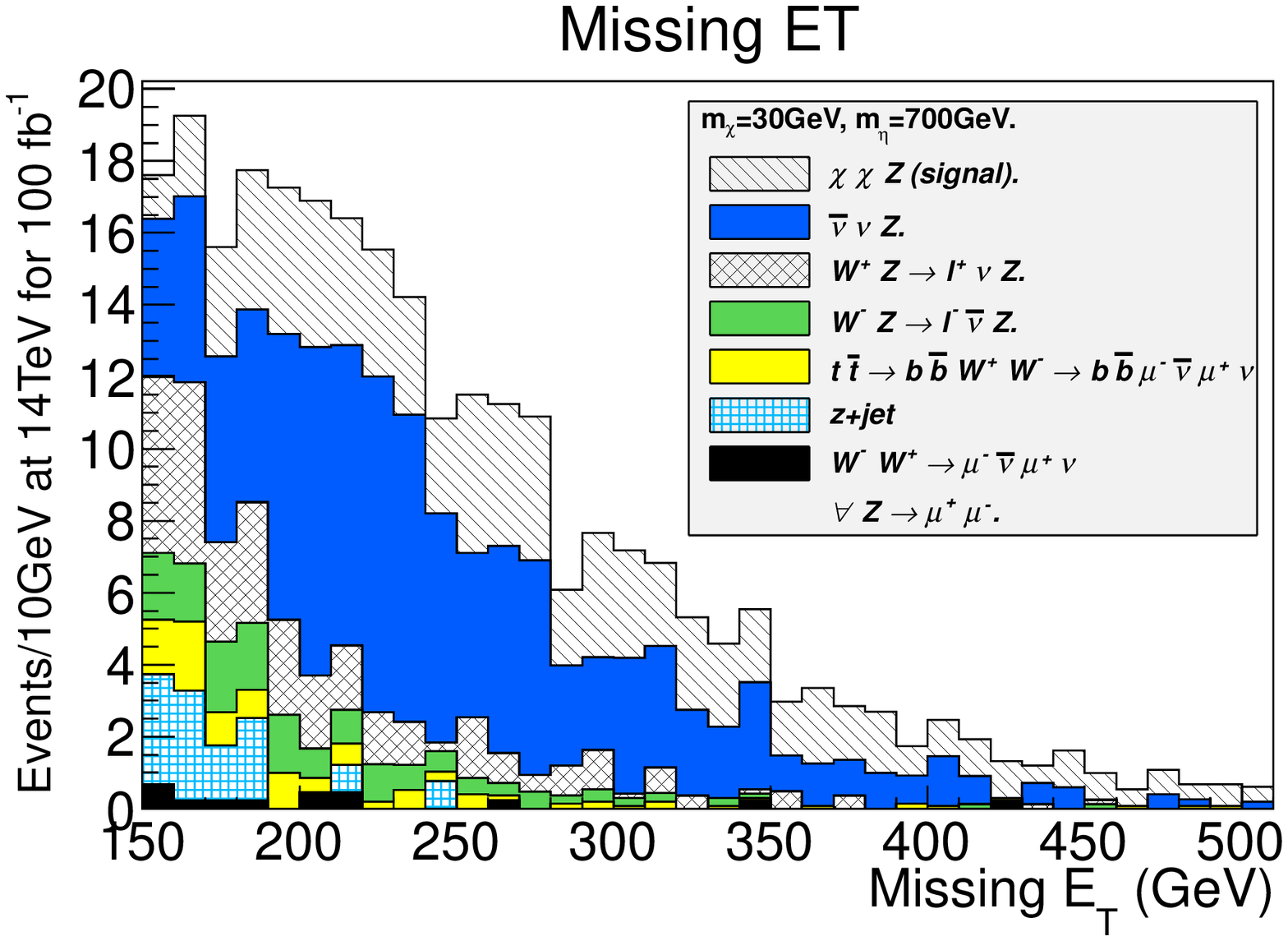}
\includegraphics[width=0.44\textwidth]{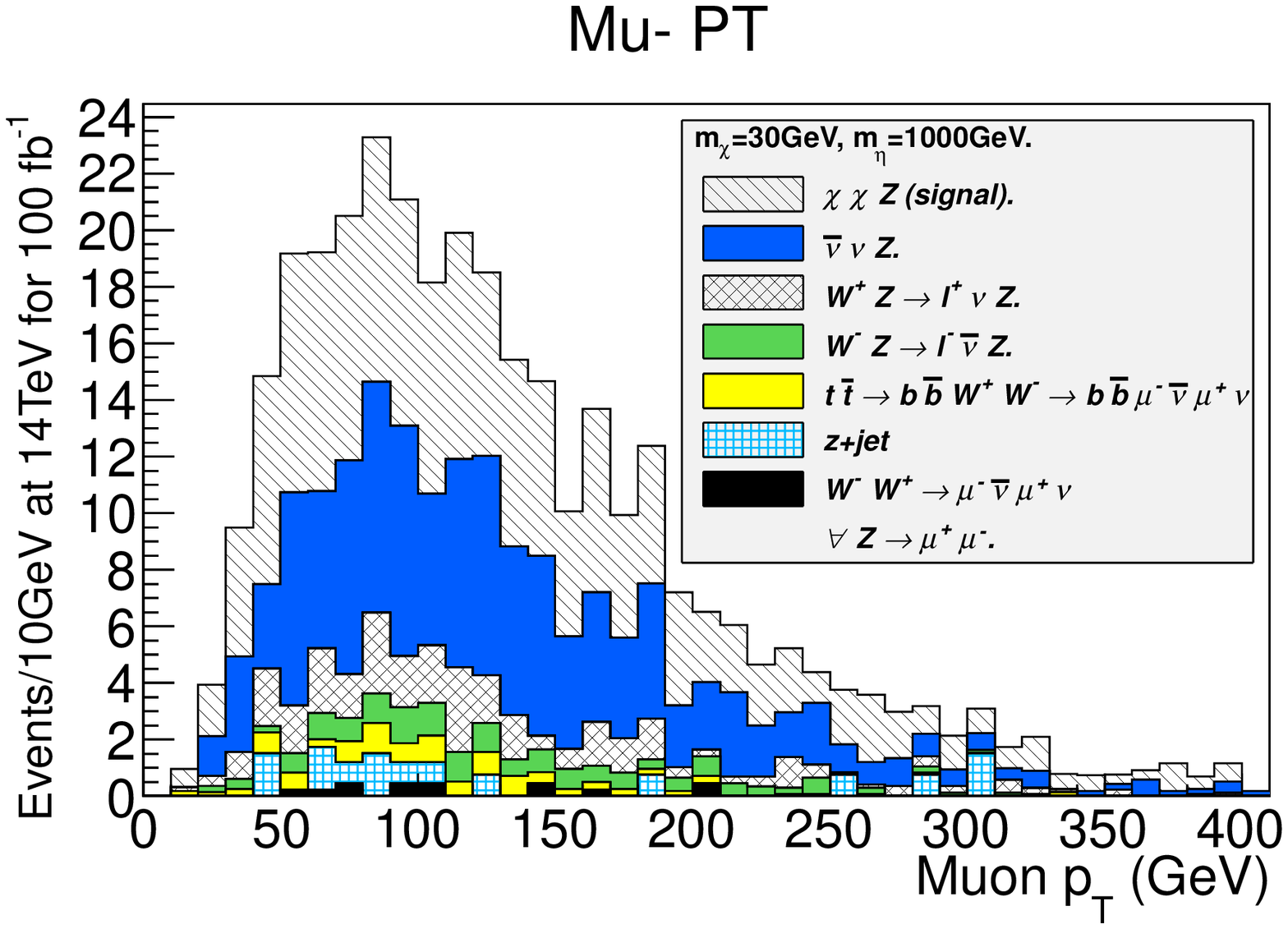}
\includegraphics[width=0.44\textwidth]{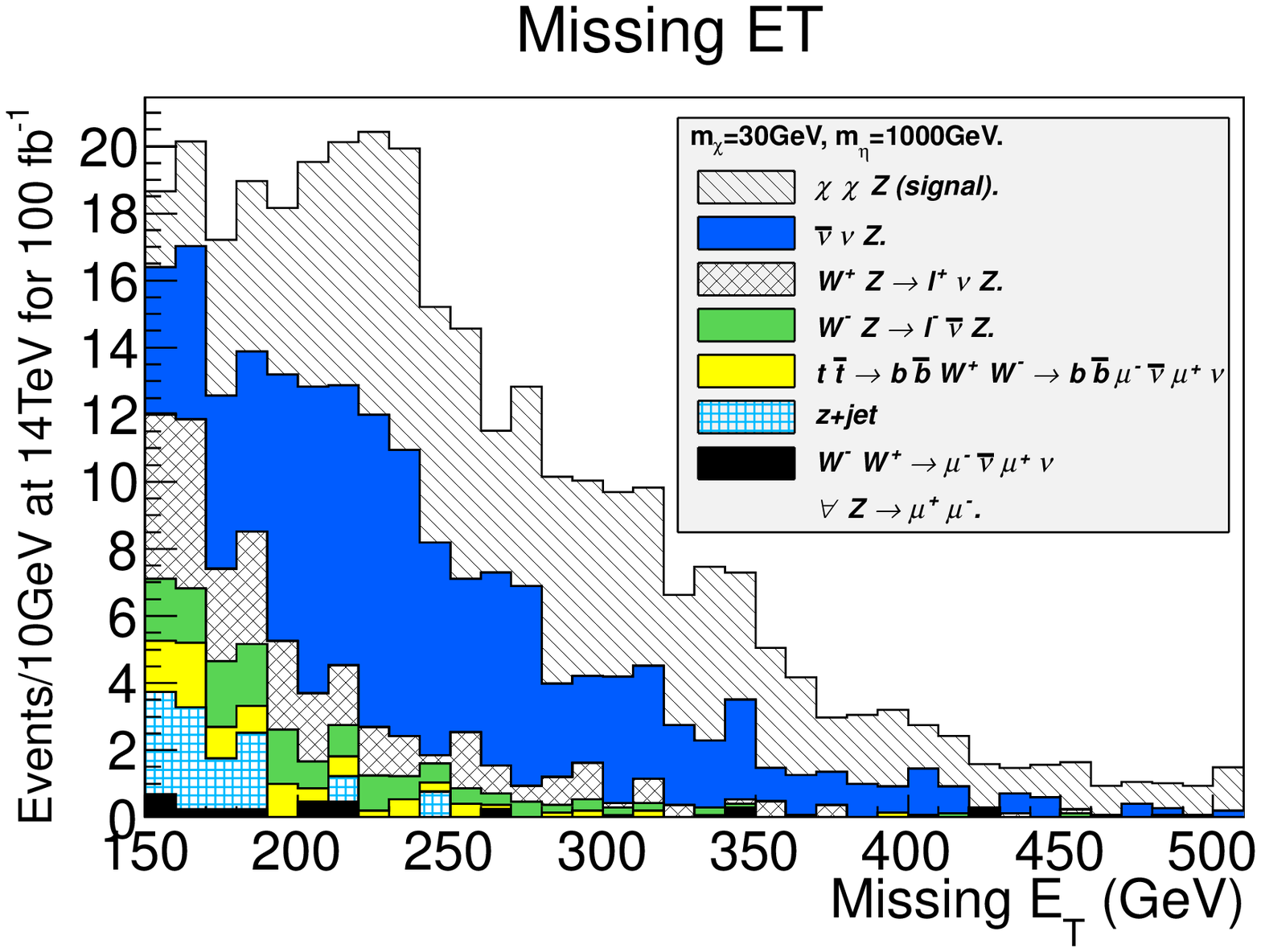}

\caption{ As Fig.~\ref{fig:results-7TeV}, with $\sqrt{s}=14$TeV and 100$\text{fb}^{-1}$ of data, for $m_\x=30$GeV, comparing $m_\eta=700$GeV and 1000GeV.\label{fig:results-meta}}
\end{figure*}

\begin{figure*}[!htbp]
\includegraphics[width=0.44\textwidth]{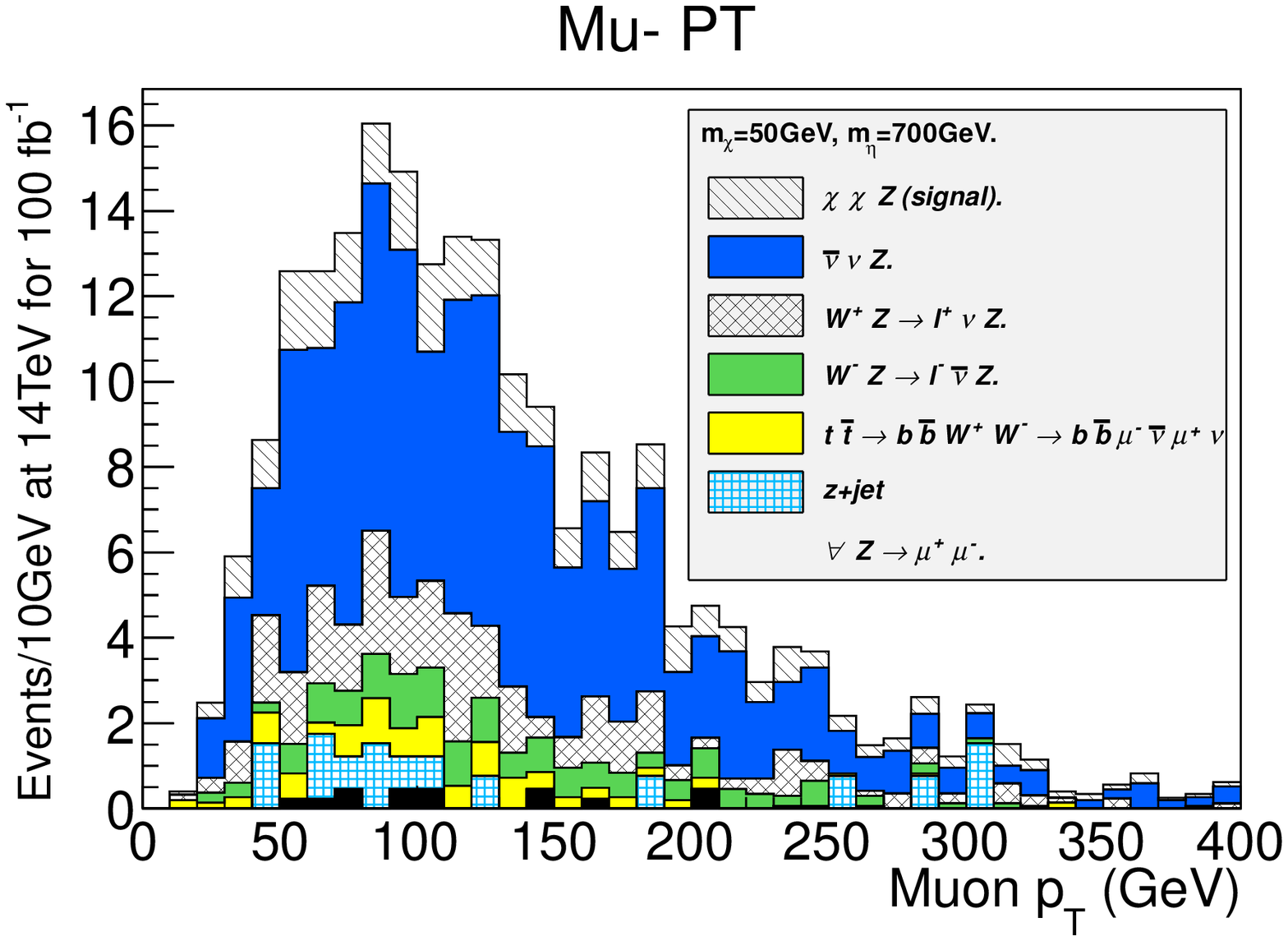}
\includegraphics[width=0.44\textwidth]{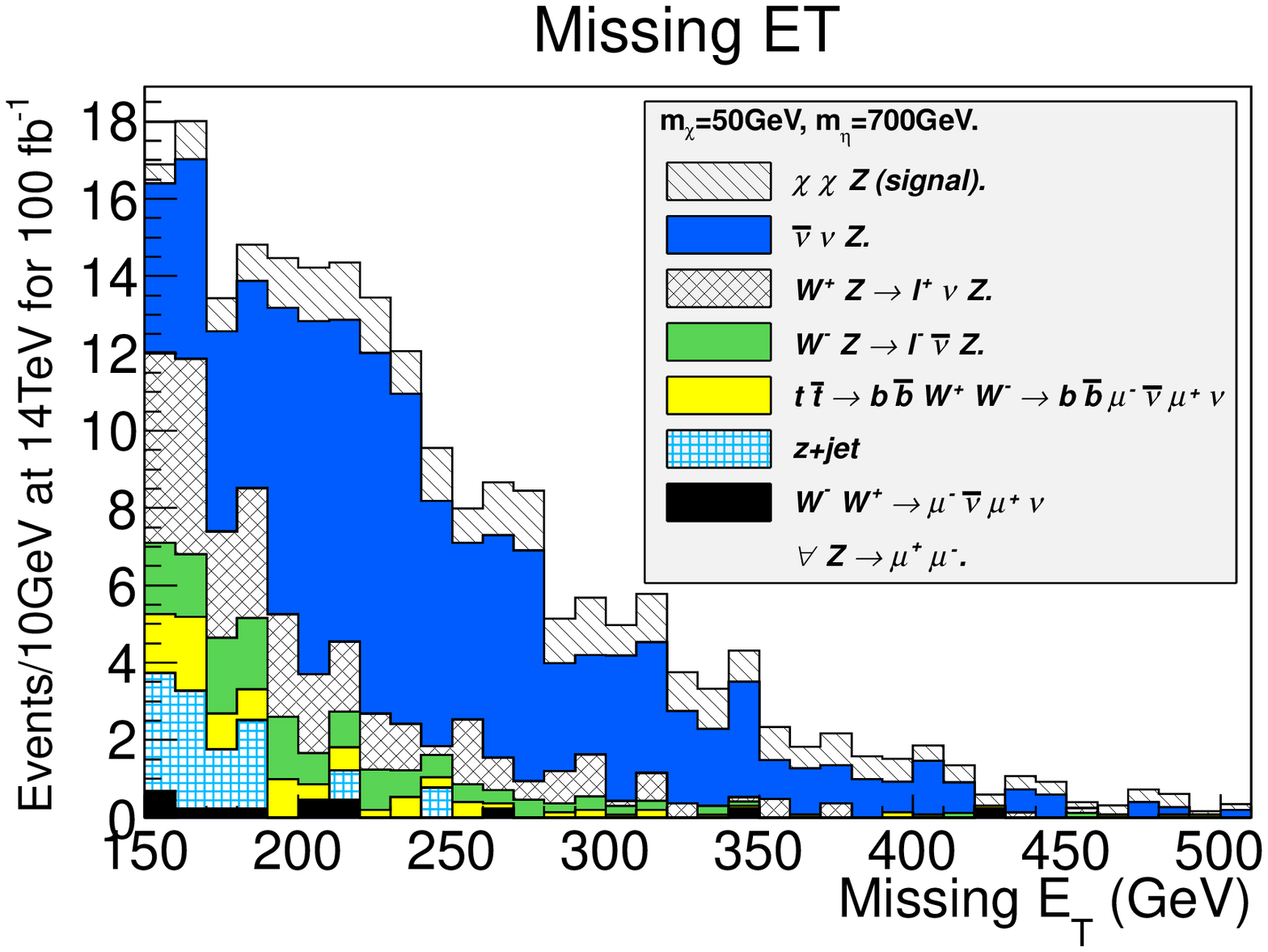}
\includegraphics[width=0.44\textwidth]{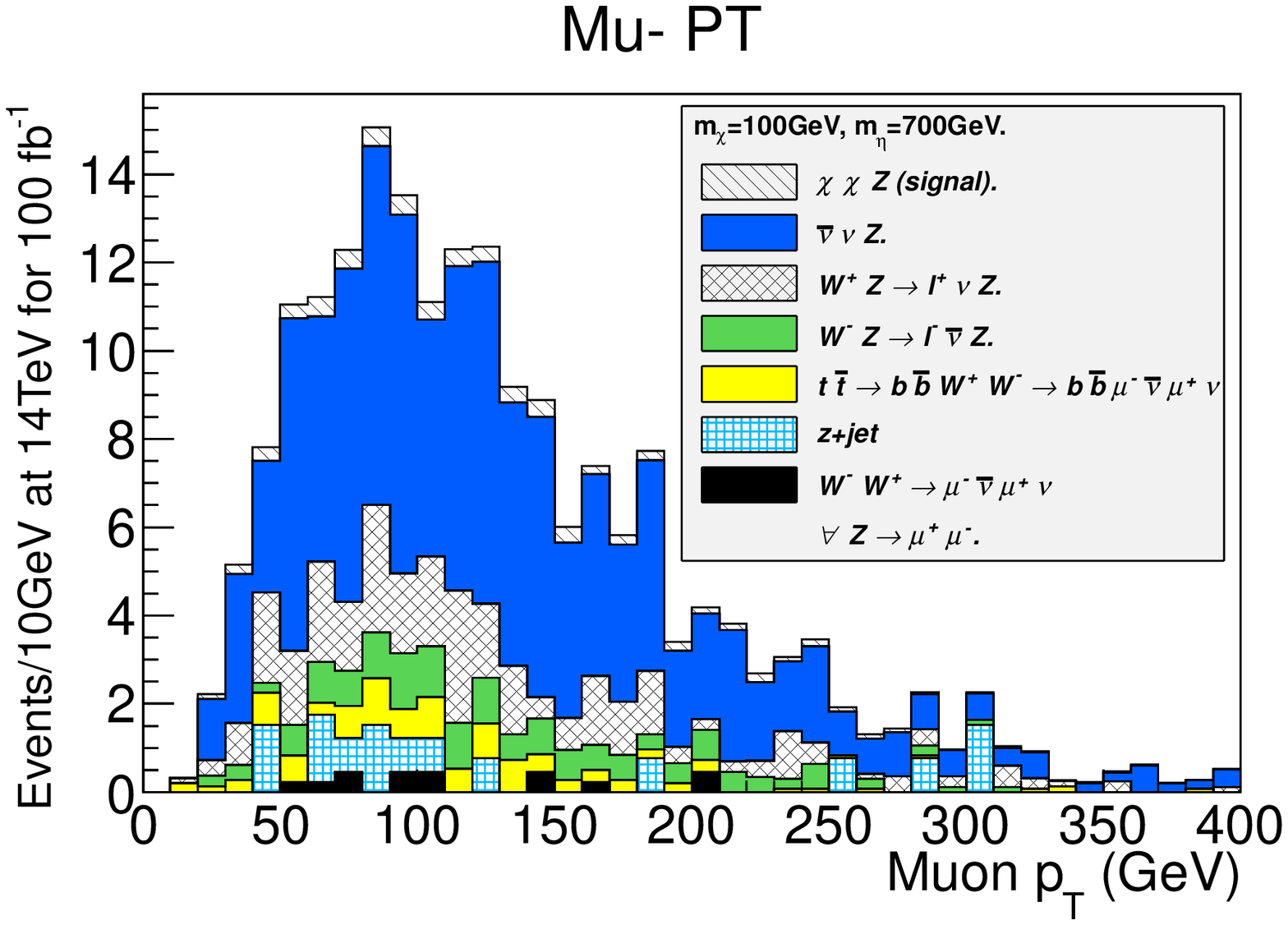}
\includegraphics[width=0.44\textwidth]{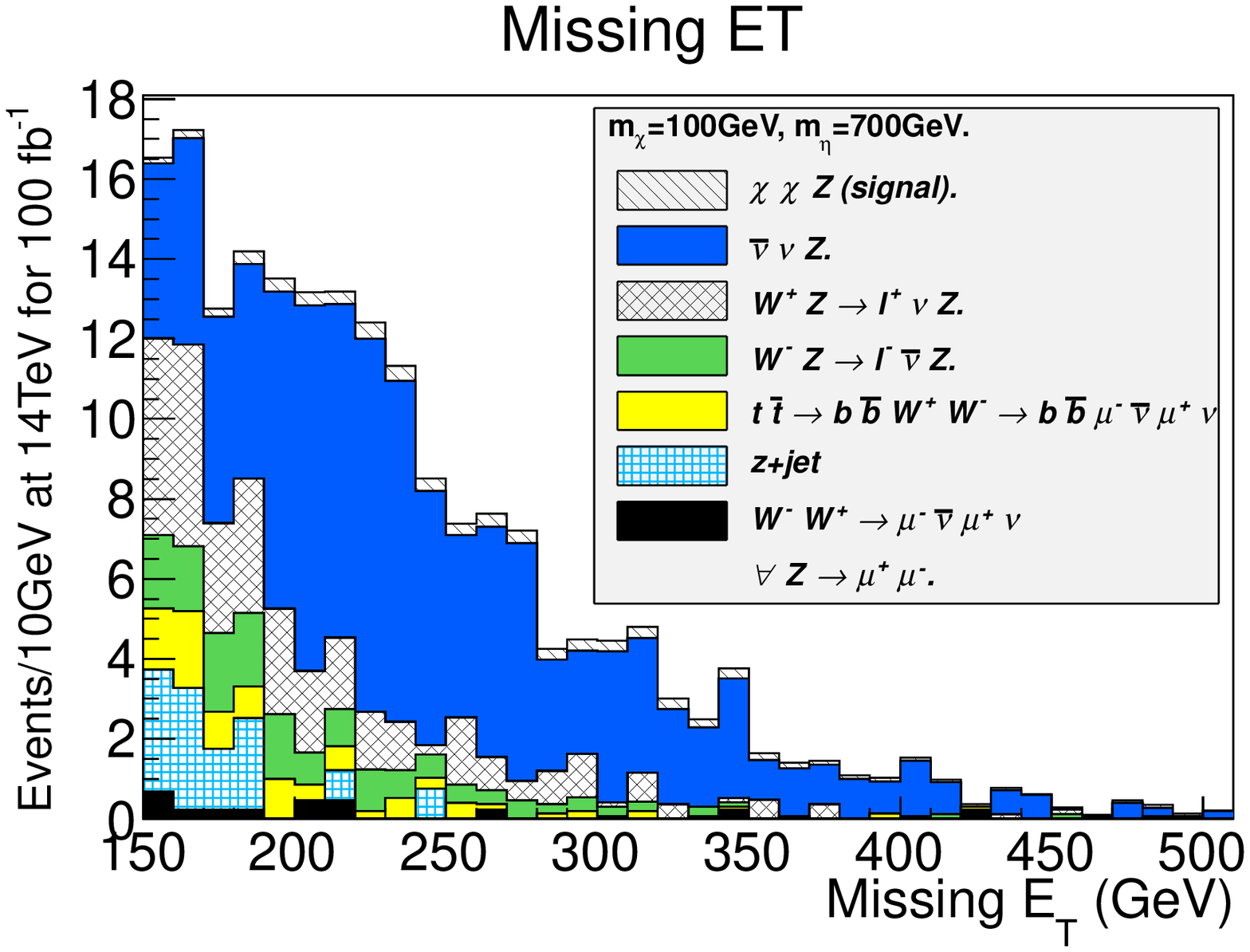}

\caption{{\bf} As Fig.~\ref{fig:results-7TeV}, with $\sqrt{s}=14$TeV and 100$\text{fb}^{-1}$ of data, for $m_\eta=700$GeV, comparing  $m_\x=$50 GeV, and 100 GeV.\label{fig:results-mchi}}
\end{figure*}

\begin{figure*}[!htbp]
\includegraphics[width=0.44\textwidth]{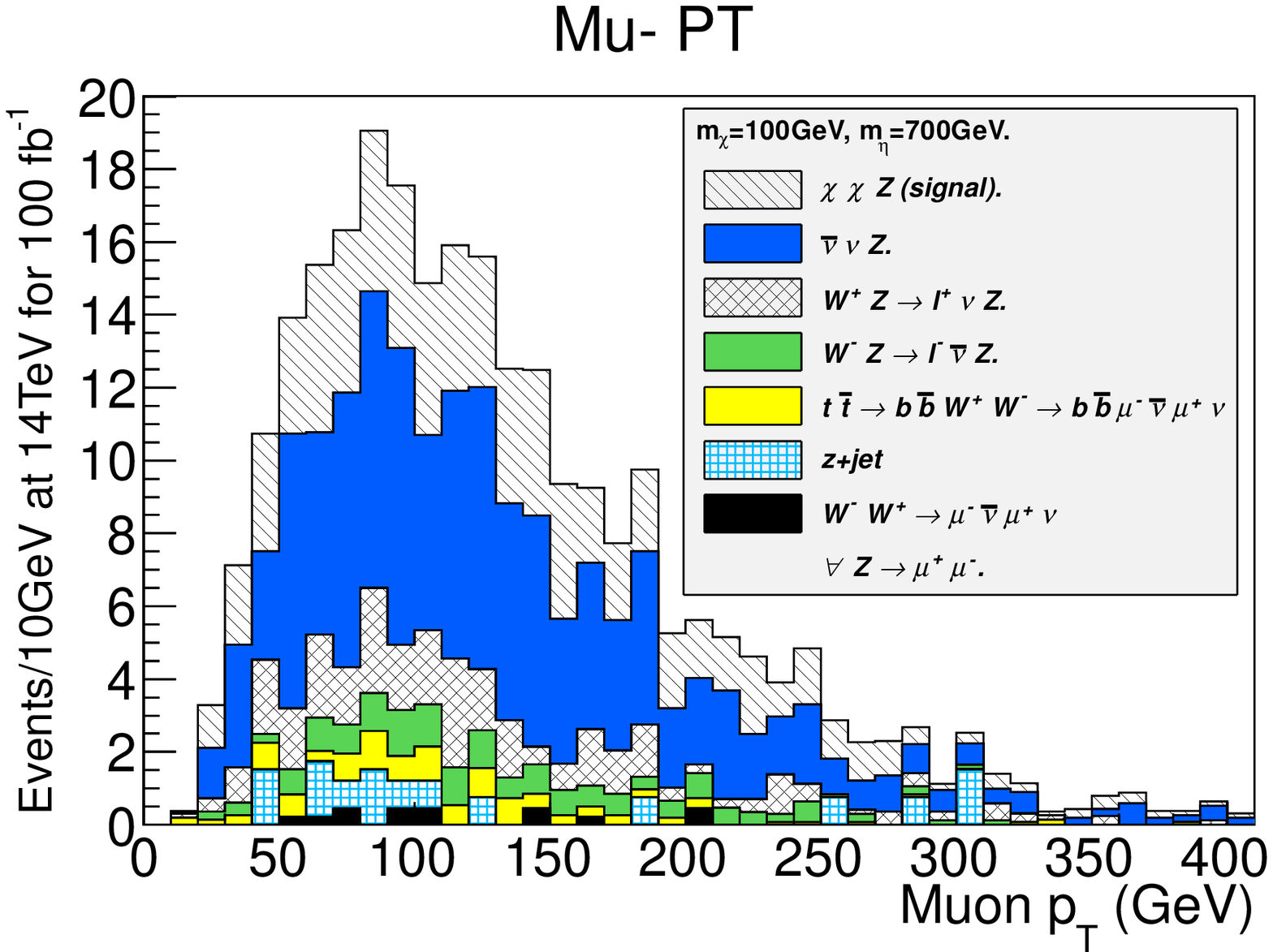}
\includegraphics[width=0.44\textwidth]{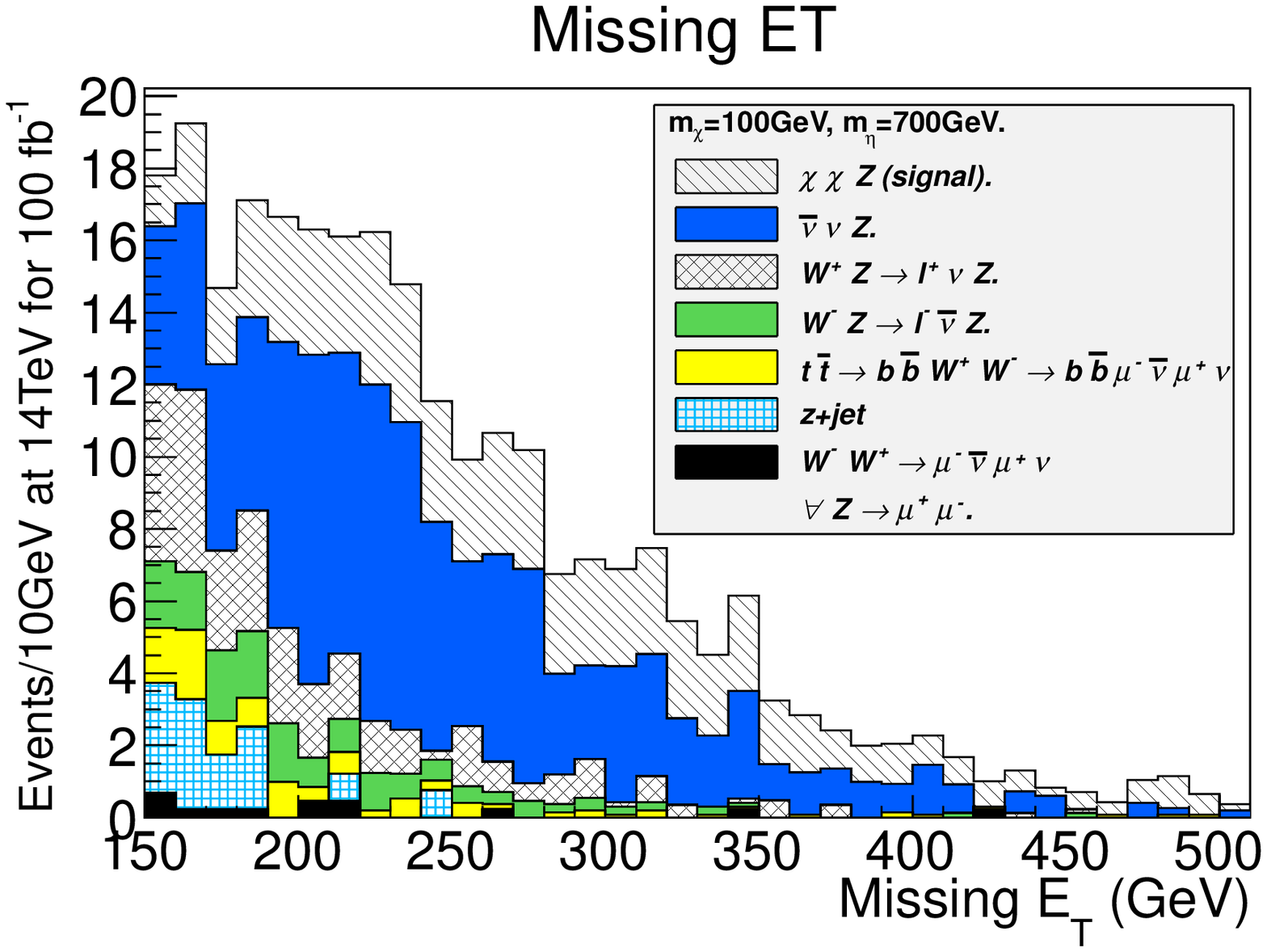}

\caption{{\bf} As Fig.~\ref{fig:results-mchi}, with $m_\x=100$GeV and $m_\eta=700$GeV, for $f_{ud}=3$, which corresponds to 10\% of present relic abundance.\label{fig:results-fud}}
\end{figure*}

The results of our simulations can be seen in
Figs.~\ref{fig:results-7TeV} -- \ref{fig:results-fud}.  These figures 
sample a range of model parameters, LHC energies and integrated
luminosities, and use values of $f_{ud}$ adherent to the constraints
in Fig.~\ref{fig:fconstraints}. Plotted are expected signal and
background events per 10GeV bin as a function of missing energy and
the $p_T$ of the $\m^{-}$, after the application of cuts outlined in
Section~\ref{subsec:EventSelection}.  The $p_T$ distributions for $\mu^{+}$
and $\mu^{-}$ differ due to the $CP$-breaking valance quark
contribution to the parton distributions of the proton.  However, given the
high CoM energy of LHC collisions, this is a small effect.  We thus do
not show the $\mu^{+}$ $p_T$ distributions, being nearly
indistinguishable from those of $\mu^-$.

In order to get accurate statistics in regions with low numbers of
events, all of our simulations except those of $Z+jet$ have at least four times as many
simulated events as expected LHC events for the given integrated luminosity.
For the $Z+jet$ background there are $\sim1.3$ simulated events per LHC event. 
All event numbers are then rescaled for our figures, 
to the number of events expected at the LHC.

All backgrounds have been significantly reduced by our choice of cuts, and
the remaining background is dominated by $Z\n\bar\n$.  This can be
understood given that it passes $Z$ selection criteria, and contains a pair of neutrinos, implying larger
amounts of $\slashed{E}_T$ than other backgrounds and a greater
resilience to our missing energy cut.

Fig.~\ref{fig:results-7TeV} shows event numbers corresponding to 
an integrated luminosity of 5$\text{fb}^{-1}$ and a CoM energy of 7TeV. 
It is clear that our selection cuts are effectively distinguishing signal from background. Results for $m_\x=10$GeV and $m_\eta=700$GeV indicate an excess of a few events after integration across all bins, demonstrating the potential for constraining the model using current data. For $m_\x=30$GeV the signal strength is significantly weaker. In this case, a CoM energy of 8TeV and an integrated luminosity of 15$\text{fb}^{-1}$ are required for the signal to be visible, as demonstrated in Fig.~\ref{fig:results-8TeV}. With the intention of studying heavier dark matter masses, we focus primarily on the higher integrated luminosities and CoM energies, for which the expected signal is significantly enhanced.

%%%%%%%%%%%%%%%%%%%%%%%%%
%%%%%%%%%%%%%%%%%%%%%%%
%%%%%%%%%%%%%%%%%%%%%%%%%
%%%%%%%%%%%%%%%%%%%%%%%

 Looking to the future, we turn our attention to the LHC's design CoM energy of 14TeV, and a larger integrated luminosity of 100$\text{fb}^{-1}$. 
Figs.~\ref{fig:results-meta} - \ref{fig:results-mchi} show the expected signal for varying model parameters $m_\x$ and $m_\eta$. 
Signal to background in most bins for the regions of parameter space considered exceed 10\%, 
reaching up to 70\% in some bins for $m_\x=30$GeV and $m_\eta=1$TeV. 
We expect such an excess, if it exists, to be clearly visible in future data.  Consequently, 
the non-observation of this signal has the potential to place strong constraints on this type of model. 

The signal strength decreases as the dark matter mass increases, approaching invisibility as $m_\x$ approaches 100GeV. Though not implemented here, this decrease in signal to background with increasing $m_\x$ could potentially be offset by optimizing cuts on measured events for a given dark matter mass. Alternatively, one could relax the requirement that $\x$ constitute all of the dark matter, allowing an increase in $f_{ud}$, which would in turn scale the production cross section. 
An example of such non-saturating dark matter can be seen in Fig.~\ref{fig:results-fud}.
Here, $f_{ud}$ is taken to be~3, in contrast to the value of 1.67 required to satisfy the relic abundance constraint. 
With $f_{ud}=3$, the $\x$ contributes only $\sim10$\% to the total dark matter abundance.

%%%%%%%%%%%%%%%%%%%%%%%%%
%%%%%%%%%%%%%%%%%%%%%%%
%%%%%%%%%%%%%%%%%%%%%%%
%%%%%%%%%%%%%%%%%%%%%%%

%%%%%%%%%%%%%%%%%%%%%%%
%%%%%%%%%%%%%%%%%%%%%%%

%%%%%%%%%%%%%%%%%%%%%%%%%%%%%%%%%%%%%%%%%%%%
\section{Conclusions}
\label{sec:conclusions}
%%%%%%%%%%%%%%%%%%%%%%%%%%%%%%%%%%%%%%%%%%%%

There are many different search channels for dark matter at the LHC,
most of those being dependent on the specifics of the model under
consideration, supersymmetric or otherwise. A key hallmark of
any of these searches are some set of high $p_T$ events, against
whose momentum large amounts of $\slashed{E}_T$ can be
reconstructed. A complete dark matter search must take into account
signatures in all possible channels. In this work, we have pointed out
the relevance of mono-$Z$ (electroweak bremsstrahlung) processes as a
search channel for dark matter. 
In particular, we advance the process $pp\rarr\x\x
Z$, where the $Z$ decays muonically. The final state appears as a pair
of high $p_T$ muons, with an invariant mass in the $Z$ window, and
large amounts of $\slashed{E}_T$.  Despite having a cross section
which is low relative to processes with jets in the final state, this
process has few Standard Model backgrounds, and these can be tamed by modest
event cuts.  By applying a specific model for dark matter, with direct dark matter 
coupling to quarks, we have demonstrated that this process in principle could be
highly visible in future data from LHC upgrades.
A dedicated study by the ATLAS or CMS collaboration with full background and detector simulation could confirm this.
We have found that for certain choices of model parameters an excess of events 
may be visible across a broad range of energy bins;
some bins contain up to a 70\% signal to background ratio.  Although the signal 
becomes weaker with rising dark matter mass, becoming negligible near $m_{\chi} \simeq 100GeV$, 
with less conservatively chosen cuts this method may prove valuable into higher dark matter
mass ranges. 
As a result, this mono-$Z$ search can at the very least provide important complementary information 
to jet and photon based dark matter searches. 
This mono-$Z$ search is relevant 
whether or not its discovery potential is competitive with monojets and mono-photons, 
due to the unique kinematical aspects of $Z$ decay in distinguishing mono-$Z$'s 
from other bremsstrahlung based search channels.

%%%%%%%%%%%%%%%%%%%%%%%%%%%%%%%%%%%%%%%%%%%
\section*{Acknkowledgements}
%%%%%%%%%%%%%%%%%%%%%%%%%%%%%%%%%%%%%%%%%%%

We thank Elisabetta Barberio, Bhaskar Dutta, Teruki Kamon, Martin White, Antonio Limosani and Nicholas Setzer for helpful discussions. 
NFB was supported by the Australian Research Council, AJG was supported by the Commonwealth of Australia, 
TJW was supported in part by U.S. DOE Award No. DE–FG05–85ER40226, and LMK, JBD, and TDJ are supported in part by the U.S. DoE grant DE-SC0008016.  LMK acknowledges the hospitality of Australian National University, and the Research School of Astronomy and Astrophysics there, where some of this work was carried out. 

\clearpage

\bibliographystyle{h-physrev5}
\bibliography{LHCpaper}

\end{document}